\documentstyle[natb_209]{mn}  

\def\dosingle#1::::{#1}  \def\dodouble#1::::{ } 

\dodouble \documentstyle[natb_209,doublespacing]{mn} ::::

\renewcommand\citep[1]{(\citealt{#1})}
\newcommand\citepf[1]{(\citealt*{#1})}    

\def\nice#1::::{#1}    \def\subm#1::::{}   

\newcommand\zzz[2]{#2}  

\def\hMpc{\mbox{h$^{-1}$ Mpc}}
\def\hkpc{\mbox{h$^{-1}$ kpc}}

\def\centreline{\centerline}
\def\.{{.}} 
\def\gtapprox{\,\lower.6ex\hbox{$\buildrel >\over \sim$} \, }
\def\ltapprox{\,\lower.6ex\hbox{$\buildrel <\over \sim$} \, }
\def\propapprox{\,\lower.6ex\hbox{$\buildrel \propto\over \sim$} \, }

\def\e{ {\scriptstyle \times} 10^}

\def\arcs{\ifmmode {'' }\else $'' $\fi}     
\def\arcm{\ifmmode {' }\else $' $\fi}       
\def\deg{\ifmmode^\circ\else$^\circ$\fi}    
\def\ttimes{{\scriptstyle \times}}

\def\fr7{7$ \hskip -0.9ex \vrule height0.8ex width0.8ex depth-0.73ex
                                                                \hskip0.1ex$}
\def\frtoday{le\space\number\day\space\ifcase\month\or
  janvier\or f\'evrier\or mars\or avril\or mai\or juin\or
  juillet\or ao\^ut\or septembre\or octobre\or novembre\or d\'ecembre\fi\space \number\year}

\def\HII{H~{\sc II}}

\def\apj{ApJ}                 
\def\aj{AJ}                       
\def\aanda{A\&A}            
\def\mnras{MNRAS}
\def\araa{AnnRevA\&A}

\nice \input epsf1990.sty ::::
\nice \def\mycaptionfont{\protect\footnotesize} ::::

\begin{document}

\newcommand\joref[5]{#1, #5, {#2, }{#3, } #4}  
\newcommand\confref[5]{#1, #5, {#2, }{#3, } #4.} 
\newcommand\inpress[5]{#1, #5, #2, in press}  
\newcommand\prepr[4]{#1, #3.  #2 (preprint)} 
\newcommand\epref[3]{#1, #3, #2}


\def\reff{$r_{\mbox{\rm \small eff}}^p$}
\def\refffifty{$r_{\mbox{\rm \small eff}}^{50\%}$}
\def\zmed{$z_{\mbox{\rm \small med}}$}
\def\rhalo{r_{\mbox{\rm \small halo}}}
\def\rhalosm{r_{\mbox{\rm \tiny halo}}}   
\def\nmean{\overline{n_{\mbox{\small est}}}}
\def\UFthhW{U_{\mbox{\small F300W}}}
\def\IFeofW{U_{\mbox{\small F814W}}}
\def\thmask{\theta_{\mbox{\small mask}}}
\def\Pgauss{P_{\mbox{\small Gauss}}}


\def\CAMK{Nicolaus Copernicus Astronomical Center, 
ul. Bartycka 18, 00-716 Warsaw, Poland}
\def\IAP{Institut d'Astrophysique de Paris, 98bis Bd Arago, F-75.014 Paris,
France}
\def\IUCAA{Inter-University Centre for Astronomy and Astrophysics, 
Post Bag 4, Ganeshkhind, Pune, 411 007, India}
\def\Stras{UMR CNRS 7550, Observatoire de Strasbourg, 
                 11 rue de l'Universit\'e, 67000 Strasbourg, France}
\def\RGO{Royal Greenwich Observatory, Madingley Road,
                 Cambridge CB3 0EZ, UK}
\def\Imperial{Astrophysics Group, Blackett Laboratory, Imperial College,
  Prince Consort Rd, London SW7~2BZ, UK}

\title{Galaxy Clustering at $z\sim2$ and Halo Radii}
\author[B.~F.~Roukema et al.]{B.~F. Roukema$^{1,2,3}$,
D.~Valls-Gabaud$^{2,4}$, B.~Mobasher$^5$ and S.~Bajtlik$^1$\\
{$^1$\CAMK}\\ {$^2$\Stras}\\ 
 {$^3$\IUCAA}\\ {$^4$\RGO}\\ {$^5$\Imperial}\\
Email: boud@iucaa.ernet.in, dvg@astro.u-strasbg.fr, b.mobasher@ic.ac.uk,
      bajtlik@camk.edu.pl}
\def\today{\frtoday}

\maketitle

\begin{abstract}
The amplitude of the angular two-point galaxy auto-correlation
function $w(\theta)$ for galaxies at $z\sim 2$ is estimated for
galaxies in the Hubble Deep Field by using
a $U<27$ complete sub-sample. 
The $U$-band selection ensures little
contamination from $z>2\.5$ galaxies, while photometric
redshifts minimise the contribution from low redshift galaxies.

(i) It is confirmed that the amplitude of the correlation can be 
corrected for the integral constraint (lack of large scale variance) 
without having to make assumptions
about the shape of the correlation function and by avoiding the
introduction of linear error terms. The estimate using
this technique is $w(\theta\approx5\arcs) =0\.10 \pm 0\.09.$
Estimators which assume a power law of a given slope and include
linear error terms would double this value.

(ii) If the biases introduced in faint galaxy selection due to 
obscuration by large
objects are not corrected for by masking areas around them, then
the estimate would be $w(\theta\approx5\arcs) =0\.16 \pm 0\.07.$

\dodouble \end{abstract} \clearpage
 \begin{abstract}  ::::  

(iii) The effective (three-dimensional) galaxy pair separation 
at 5{\arcs} and this redshift range is $\approx 25${\hkpc}$-250${\hkpc}, 
so the correction to the spatial correlation function $\xi(r)$ 
due to exclusion of overlapping 
galaxy dark matter haloes should be considered. 
For clustering stable in proper units
in an $\Omega=1,\lambda=0$ universe, 
our $w(5\arcs)$ estimate 
(a) implies a present-day correlation length of 
$r_0\sim2\.6^{+1.1}_{-1.7}${\hMpc} if halo 
overlapping is ignored, but (b) for 
a present-day correlation length of $r_0=5\.5${\hMpc} 
implies that a typical halo exclusion radius is
$\rhalo=70^{+420}_{-30}$\hkpc. 
For $\Omega_0=0\.1,\lambda_0=0\.9,$ the corresponding values 
are (a)  $r_0\sim5\.8^{+2.4}_{-3.9}${\hMpc} and 
(b) $\rhalo< 210${\hkpc} ($1\sigma$ upper limit). 

(iv) The decreasing correlation period (DCP) of a high initial bias
in the spatial correlation function is not detected at this redshift.
For an $\Omega=1,\lambda=0$ universe and stable clustering in proper 
units, possible detections of the DCP in other work would
imply that the values of $\xi$ at redshifts greater than 
$z_t = 1\.7\pm0\.9$ would be $[(1+z)/(1+z_t)]^{2.1\pm3.6}$ times their 
values at $z_t,$ which is consistent with
our lack of a detection at $z\sim2.$
\end{abstract}



\def\fwimage{
\begin{figure*}
\centering 
\nice \centreline{\epsfxsize=13cm
\zzz{\epsfbox[41 36 570 563]{"`gunzip -c wimage.ps.gz"} }
{\epsfbox[41 36 570 563]{"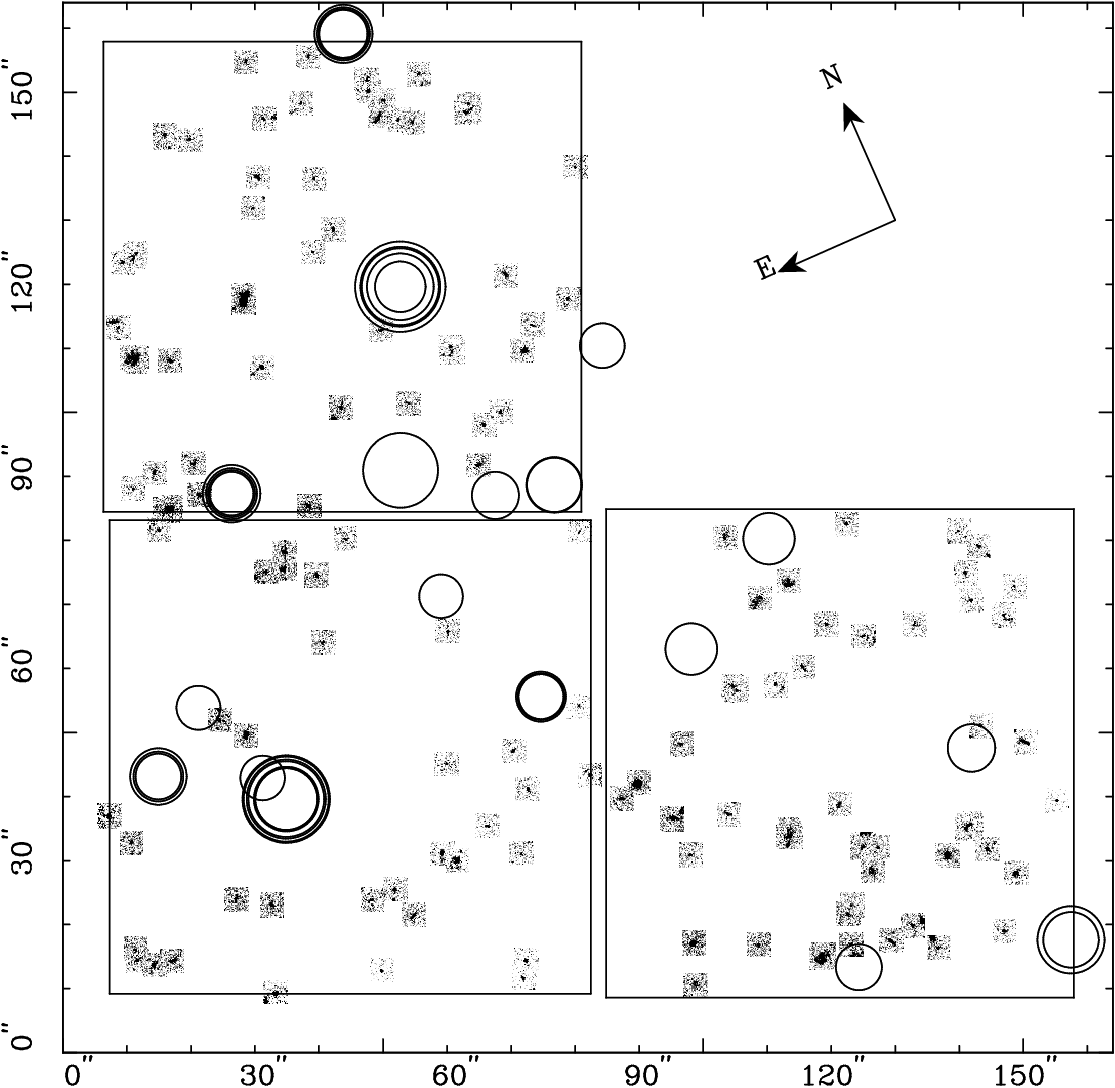"} }
}
::::
\caption{ \mycaptionfont
\label{f-wimage}
HDF $U$-selected $z\sim2$ galaxies shown as $4\arcs \times 4\arcs$ images 
extracted from the $1024\times 1024$ pixel F300W mosaic of the
HDF archive. The pixel scale is about $6\.25$ pix/{\arcs}. North is to
the upper left, east is to the lower left. 
Circles show areas excluded due to objects occupying
large solid angles from which the $z\approx2$ galaxies would not
be selected in an unbiased way. 
Concentric circles indicate that \protect\citet{Will96} chose various
options for uniting bright `split' galaxies; in such cases, 
the largest of the concentric circles is the effective exclusion zone here.
The boundaries used by
\protect\cite{Mob98} are shown for the three WFC fields, labelled
\#2 (upper left), \#3 (bottom left) and \#4 (bottom right).
One excluded area (circle) appears in the Planetary Camera field,
which is not analysed.
}
\end{figure*} 
} 

\def\fwimagenohol{
\begin{figure*}
\centering 
\nice \centreline{\epsfxsize=13cm
\zzz{\epsfbox[41 36 570 563]{"`gunzip -c imnhol.ps.gz"} }
{\epsfbox[41 36 570 563]{"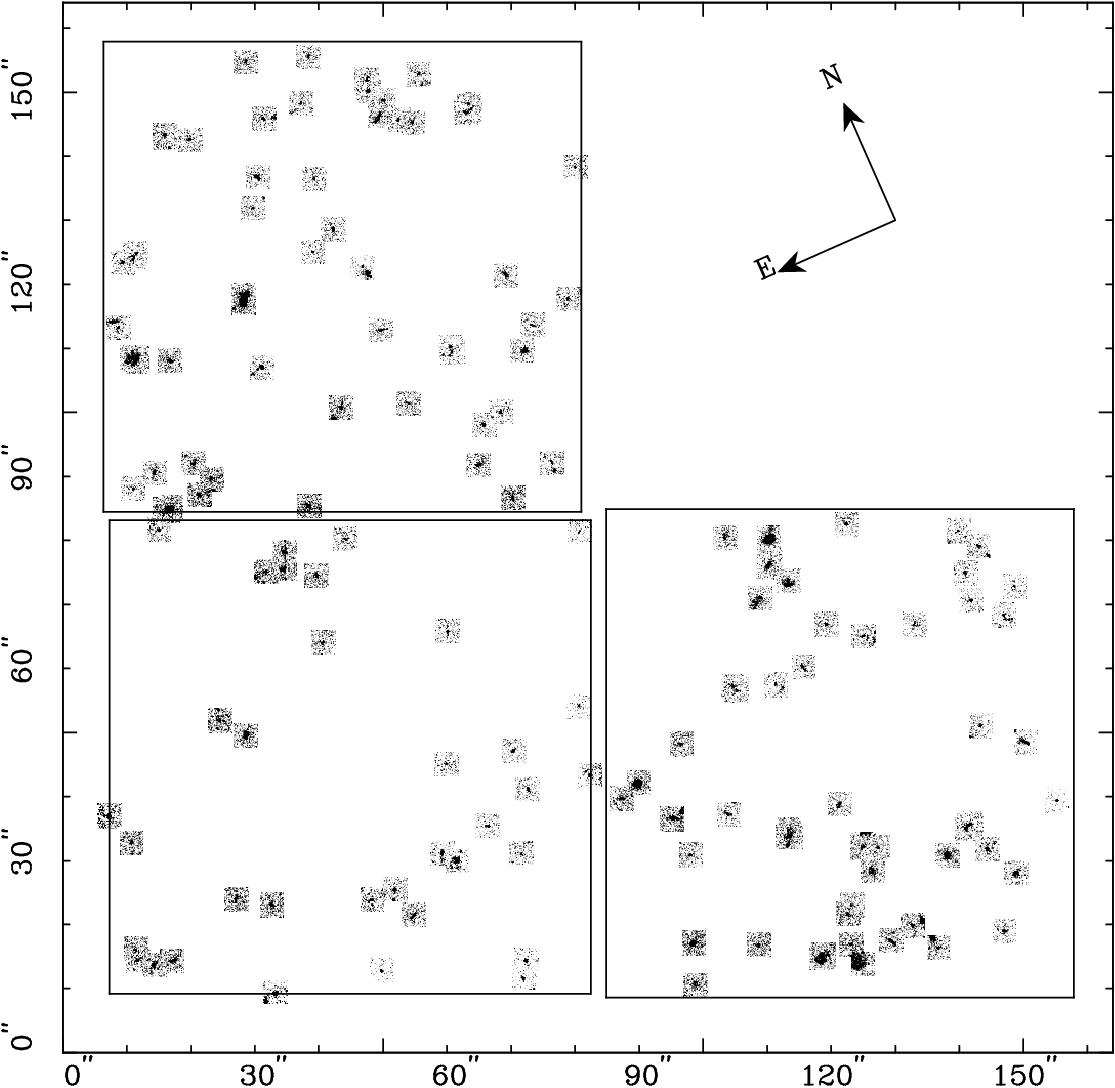"} }
}
::::
\caption{ \mycaptionfont
\label{f-wimage_nhol}
HDF $U$-selected $z \sim2$ galaxies, as for 
Fig.~\protect\ref{f-wimage} but without any masking of areas 
around large objects.
}
\end{figure*} 
} 

\def\fwiformtwo{ 
\begin{figure}
\centering 
\nice \centreline{\epsfxsize=8cm
\zzz{\epsfbox[54 41 531 528]{"`gunzip -c w2____.ps.gz"} }
{\epsfbox[54 41 531 528]{"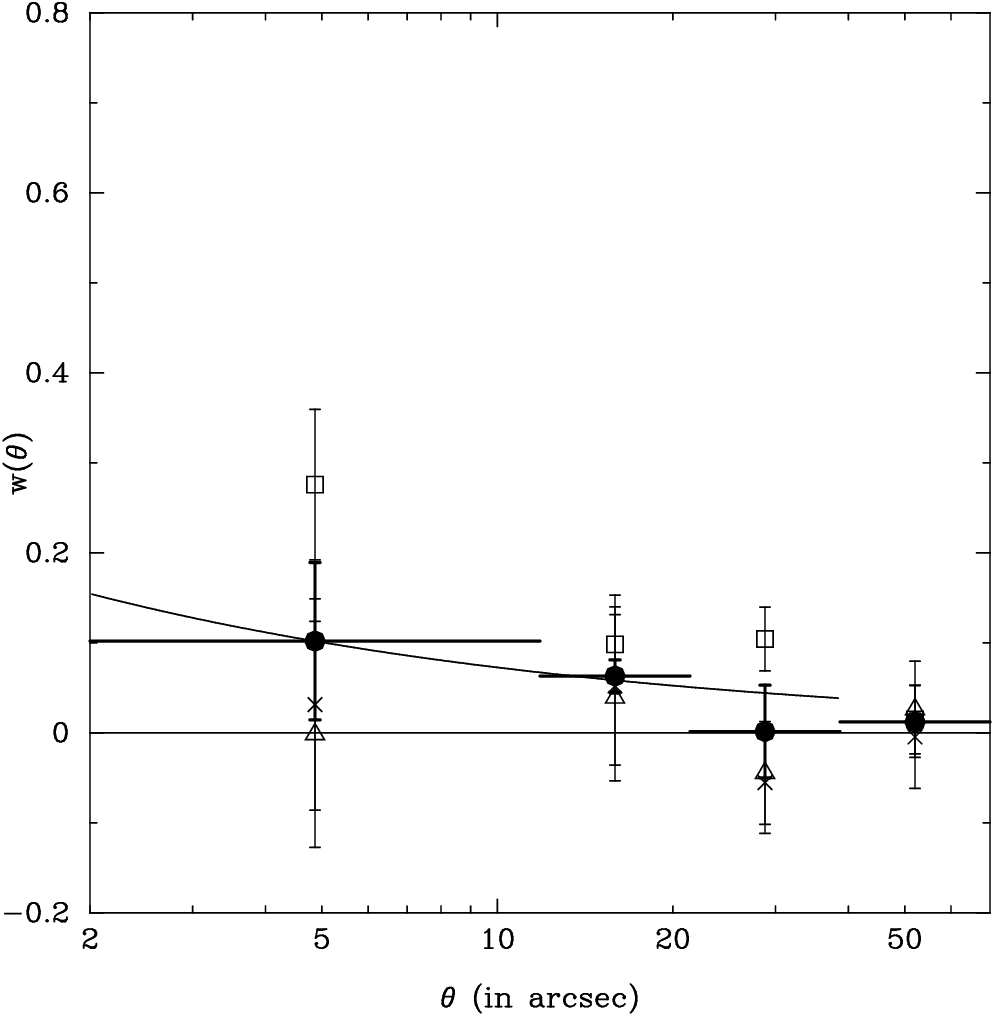"} }
}
::::
\caption{ \mycaptionfont
\label{f-wiform2}
Angular correlation function $w$ shown as a function of angle 
$\theta$ for the three WFC fields, shown as crosses (chip \#2), squares 
(chip \#3) and triangles (chip \#4), using eq.~(24) of 
\protect\citet{Ham93} to estimate $w,$ where $\nmean$ is 
the mean number density of the fields (eq.~\protect\ref{e-wham24} here)
and where regions occupied by large objects are masked.
Error bars on these points are Poissonian. The filled circles and their vertical 
error bars (in bold) show the mean values of $w$ 
for the three chips considered as independent samples and the standard
errors in the mean. Horizontal error bars show the bin sizes. 
A power law fit to the mean values is shown. 
See Table~\protect\ref{t-wamp} 
for key numerical values in this and following figures.
}
\end{figure} 
} 

\def\fwiformtwonohol{ 
\begin{figure}
\centering 
\nice \centreline{\epsfxsize=8cm
\zzz{\epsfbox[54 41 531 528]{"`gunzip -c w2F___.ps.gz"} }
{\epsfbox[54 41 531 528]{"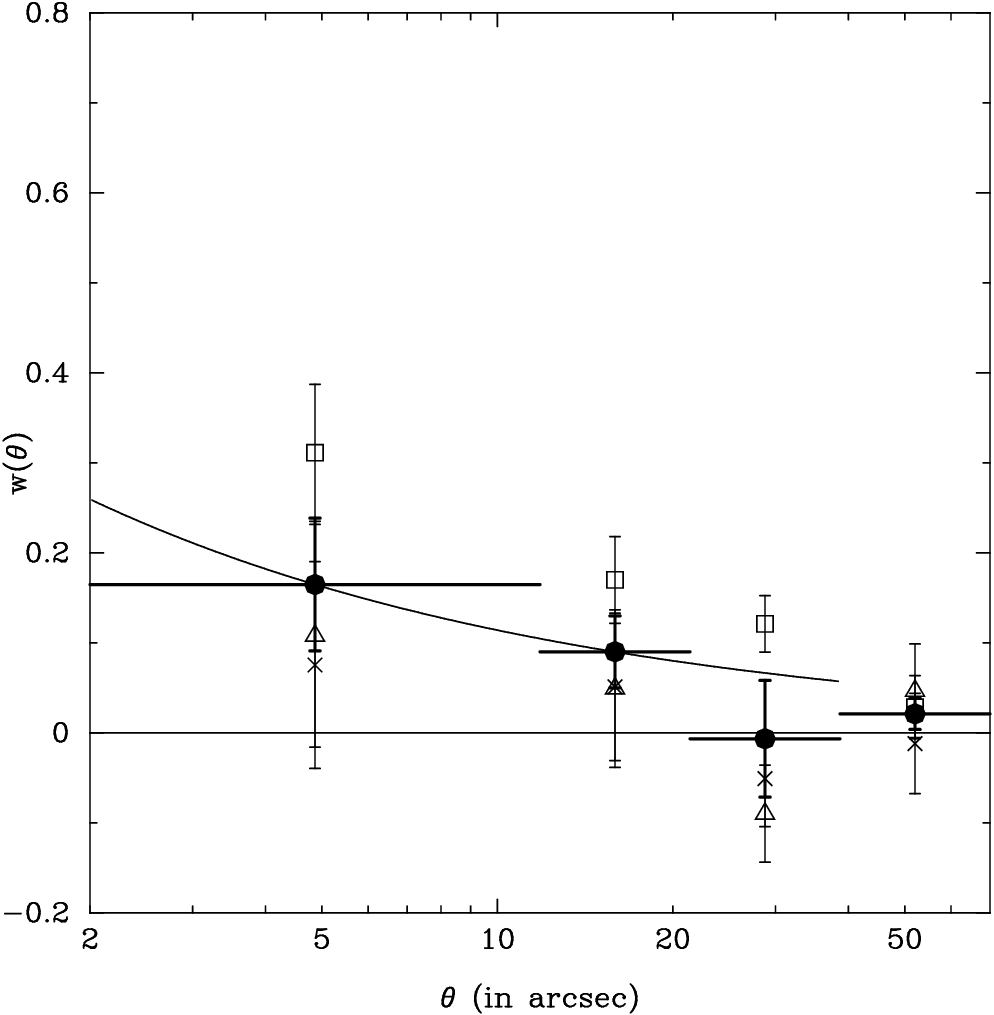"} }
}
::::
\caption{ \mycaptionfont
\label{f-wiform2nh}
As for Fig.~\protect\ref{f-wiform2}, but without any masking for
regions around bright objects.
}
\end{figure} 
} 

\def\fwiformtwonomg{ 
\begin{figure}
\centering 
\nice \centreline{\epsfxsize=8cm
\zzz{\epsfbox[54 41 531 528]{"`gunzip -c w2_F__.ps.gz"} }
{\epsfbox[54 41 531 528]{"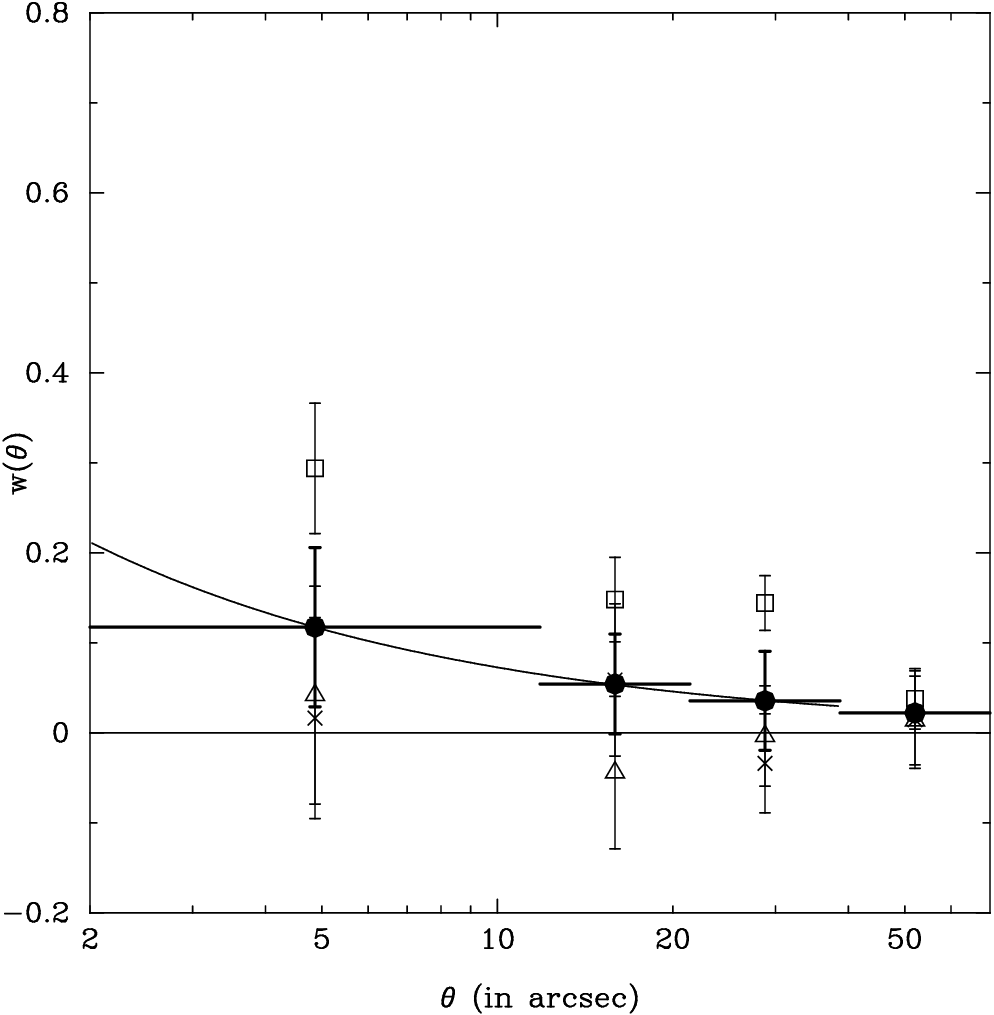"} }
}
::::
\caption{ \mycaptionfont
\label{f-wiform2nm}
As for Fig.~\protect\ref{f-wiform2}, but without any merging
of close objects.
}
\end{figure} 
} 

\def\fwiformtwofit{ 
\begin{figure}
\centering 
\nice \centreline{\epsfxsize=8cm
\zzz{\epsfbox[54 41 531 528]{"`gunzip -c w4____.ps.gz"} }
{\epsfbox[54 41 531 528]{"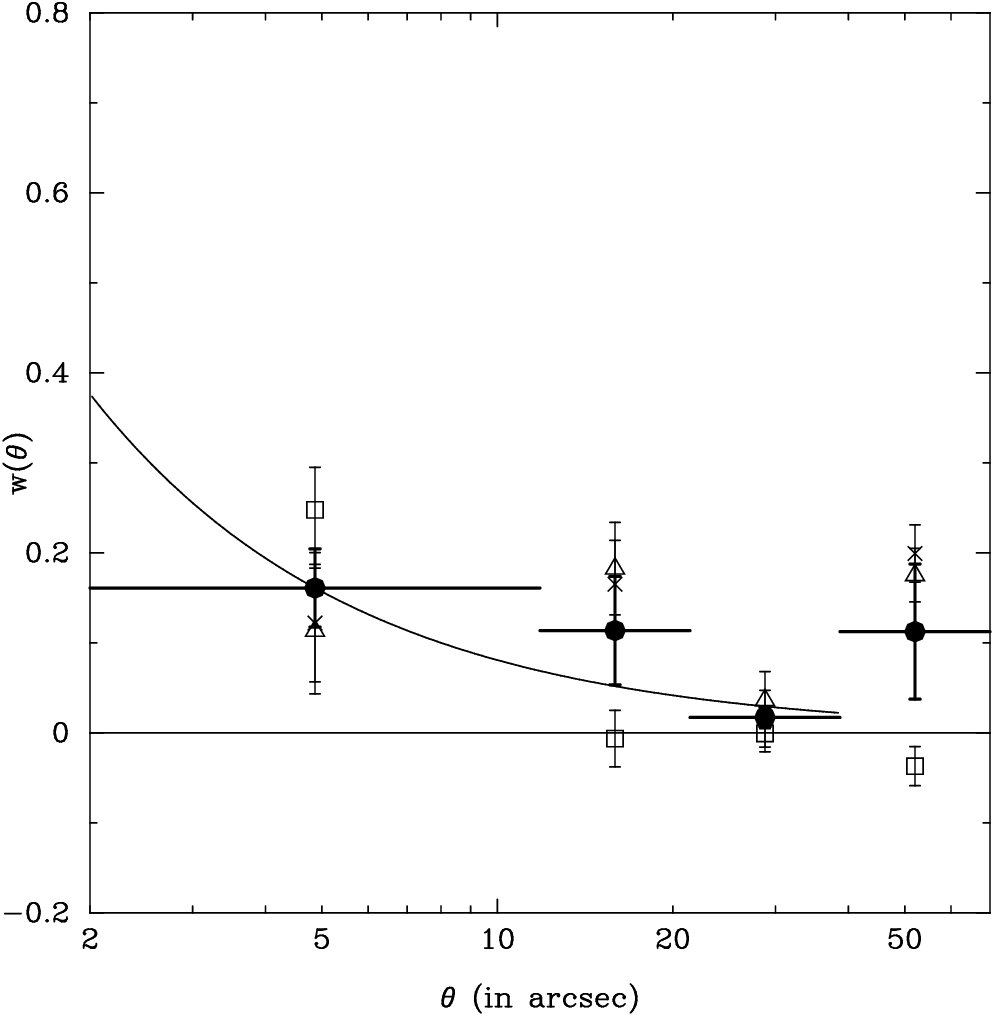"} }
}
::::
\caption{ \mycaptionfont
\label{f-wiform2fit}
As for Fig.~\protect\ref{f-wiform2}, but varying $\nmean$ in order
to best fit $w$ to a power law with slope $1-\gamma=0\.8.$ 
The slope of the curve shown is $1-\gamma=0\.85.$
}
\end{figure} 
} 

\def\fwiformthrA{ 
\begin{figure}
\centering 
\nice \centreline{\epsfxsize=8cm
\zzz{\epsfbox[54 41 531 528]{"`gunzip -c w3__F_.ps.gz"} }  
{\epsfbox[54 41 531 528]{"w3__F_.ps"} }
}
::::
\caption{ \mycaptionfont
\label{f-wiform3a}
As for Fig.~\protect\ref{f-wiform2}, but the calculation of 
$w$ is based on \protect\citeauthor{LS93}'s (1993) estimator. 
A power law cannot be fit to the mean values, since only one
is positive.
}
\end{figure} 
} 

\def\fwiformthrB{ 
\begin{figure}
\centering 
\nice \centreline{\epsfxsize=8cm
\zzz{\epsfbox[54 41 531 528]{"`gunzip -c w3____.ps.gz"} }
{\epsfbox[54 41 531 528]{"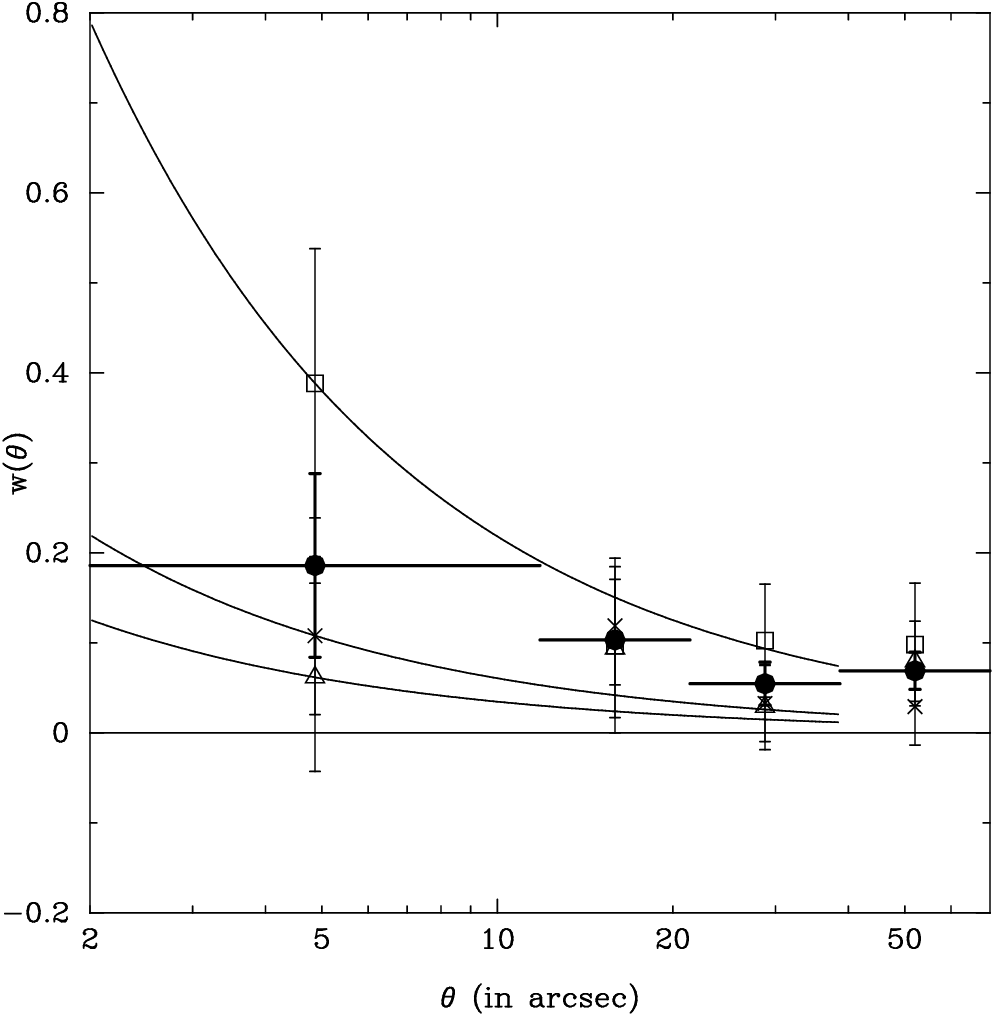"} }
}
::::
\caption{ \mycaptionfont
\label{f-wiform3b}
As for Fig.~\protect\ref{f-wiform2}, except the calculation of 
$w$ is based on \protect\citeauthor{LS93}'s (1993) estimator
plus an integral constraint additive correction 
(eq.~\protect\ref{e-ls93ic}) for 
individual fields
such that the corrected correlations are power laws of slope 
$1-\gamma=-0\.8$ (cf. \protect\citealt{Vill97,Conn98}). 
The power laws required for the corrections for the individual
fields are shown.
The mean values are calculated after the corrections.
}
\end{figure} 
} 

\def\fwiformone{ 
\begin{figure}
\centering 
\nice \centreline{\epsfxsize=8cm
\zzz{\epsfbox[54 41 531 528]{"`gunzip -c w1__F_.ps.gz"} }
{\epsfbox[54 41 531 528]{"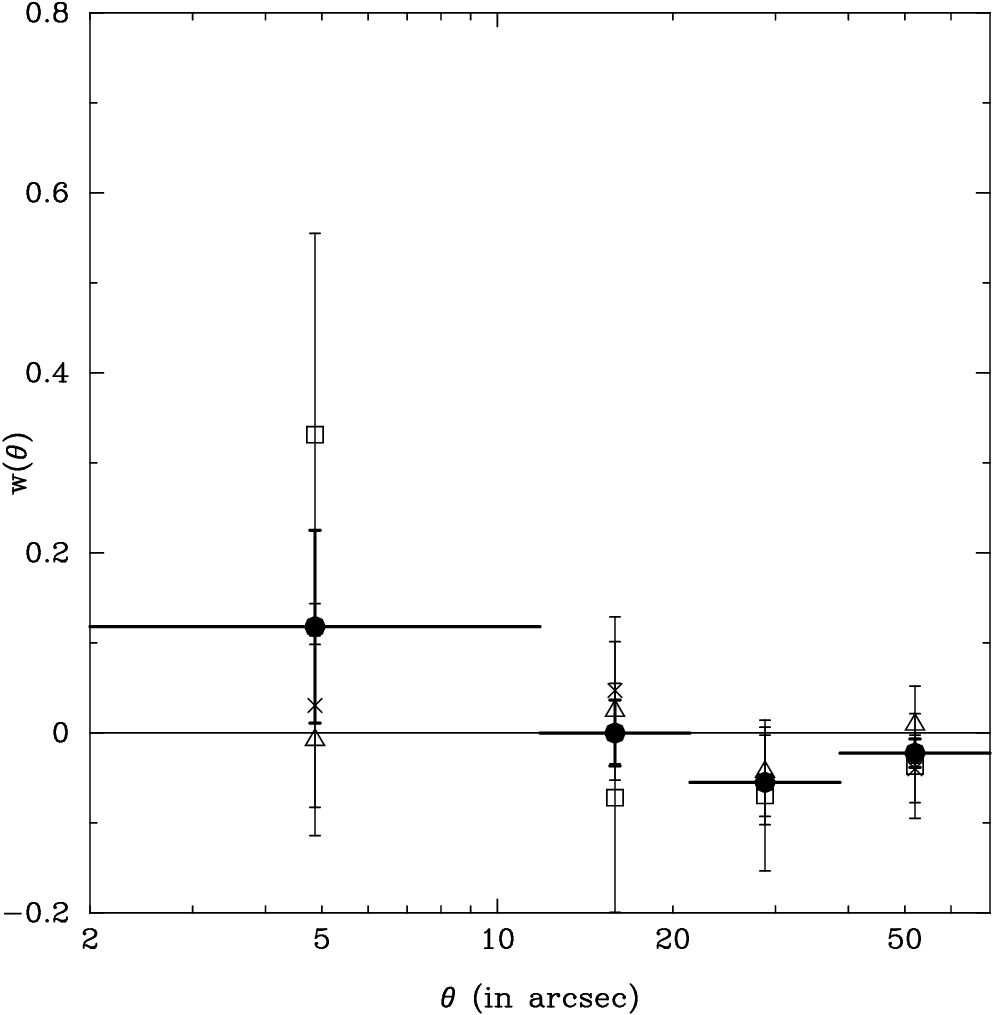"} }
}
::::
\caption{ \mycaptionfont
\label{f-wiform1}
As for Fig.~\protect\ref{f-wiform2}, 
but using eq.~(15) of \protect\citet{Ham93} 
[eq.~(\protect\ref{e-wham15}) here].
}
\end{figure} 
} 

\def\fzdis{ 
\begin{figure}
\centering 
\nice \centreline{\epsfxsize=8cm
\zzz{\epsfbox[54 44 458 379]{"`gunzip -c zdisU2.ps.gz"} }
{\epsfbox[54 44 458 379]{"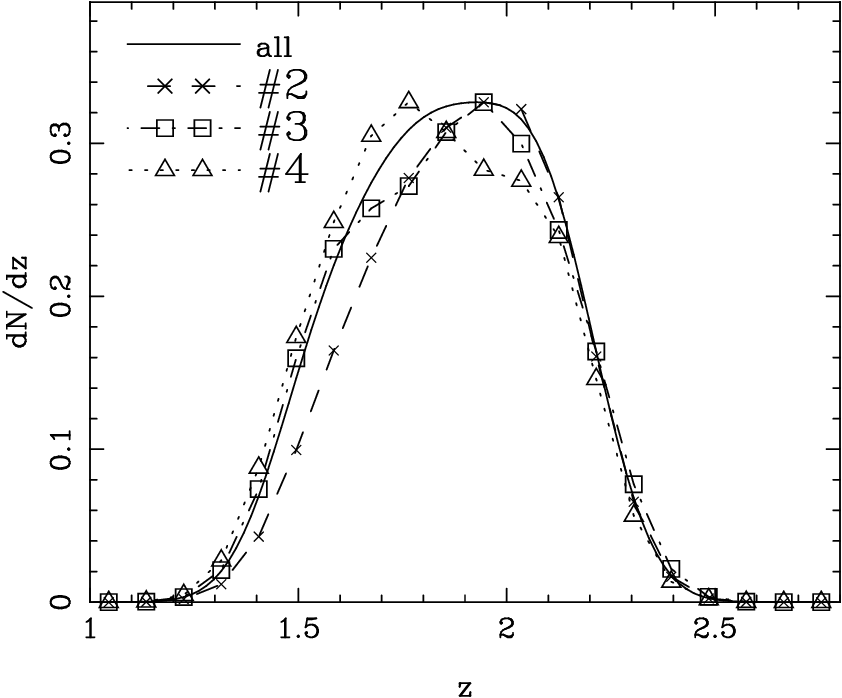"} }
}
::::
\caption{ \mycaptionfont
\label{f-zdis}
The photometric redshift distribution of the $1\.5 < z < 2\.5$ 
$U$-selected HDF galaxies for the individual WFC fields and 
in total (normalised).
}
\end{figure}
}

\def\fwrzero{ 
\begin{figure}
\centering 
\nice \centreline{\epsfxsize=8cm
\zzz{\epsfbox[50 34 523 451]{"`gunzip -c wr0.ps.gz"} }
{\epsfbox[50 34 523 451]{"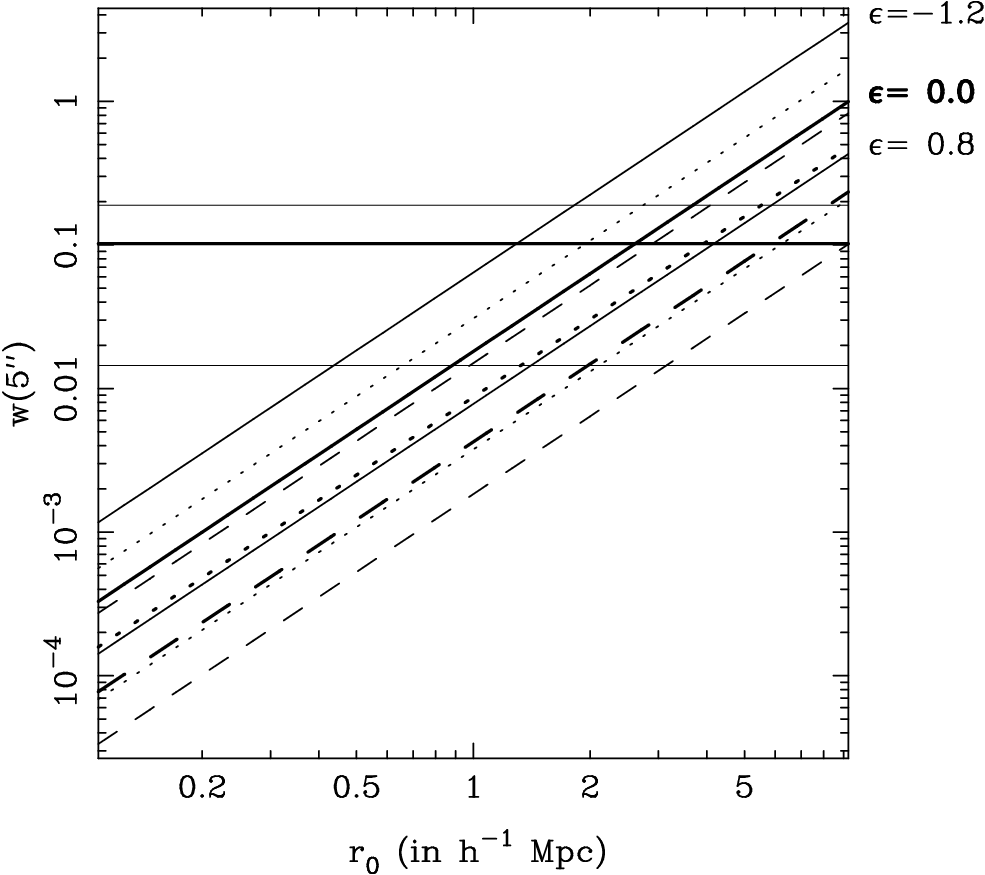"} }
}
::::
\caption{ \mycaptionfont
\label{f-wr0}
Angular correlation function amplitude $w(5\arcs)$ calculated using
eq.~(\protect\ref{e-limber}) and 
eq.~(\protect\ref{e-xieps}) for a cutoff in the spatial correlation 
function $\xi$ at pair separations $\ltapprox r_0$ 
[eq.~(\protect\ref{e-xieps})], shown against $r_0$ in \hMpc.
The three sets of curves from top to bottom are for 
correlation evolution parameters $\epsilon=-1\.2, 0\.0$ and $0\.8$
respectively. Solid, dotted and dashed curves are for metric parameters
($\Omega_0=1, \lambda_0=0$),
($\Omega_0=0\.1, \lambda_0=0$) and
($\Omega_0=0\.1, \lambda_0=0\.9$) respectively.
The photometric redshift distribution of the $1\.5 < z < 2\.5$ 
$U$-selected HDF galaxies (Fig.~\protect\ref{f-zdis}) 
is used in the integrals of eq.~(\protect\ref{e-limber}).
The HDF estimate of $w(5\arcs)$ obtained using eq.~(24) of 
\protect\cite{Ham93} (Fig.~\protect\ref{f-wiform2})
is reproduced here by the horizontal lines, mean in bold, error bar
as thin lines. $1-\gamma=-0\.8$ is adopted.
}
\end{figure} 
} 

\def\fwrhalo{ 
\begin{figure}
\centering 
\nice \centreline{\epsfxsize=8cm
\zzz{\epsfbox[50 34 523 451]{"`gunzip -c wrhalo.ps.gz"} }
{\epsfbox[50 34 523 451]{"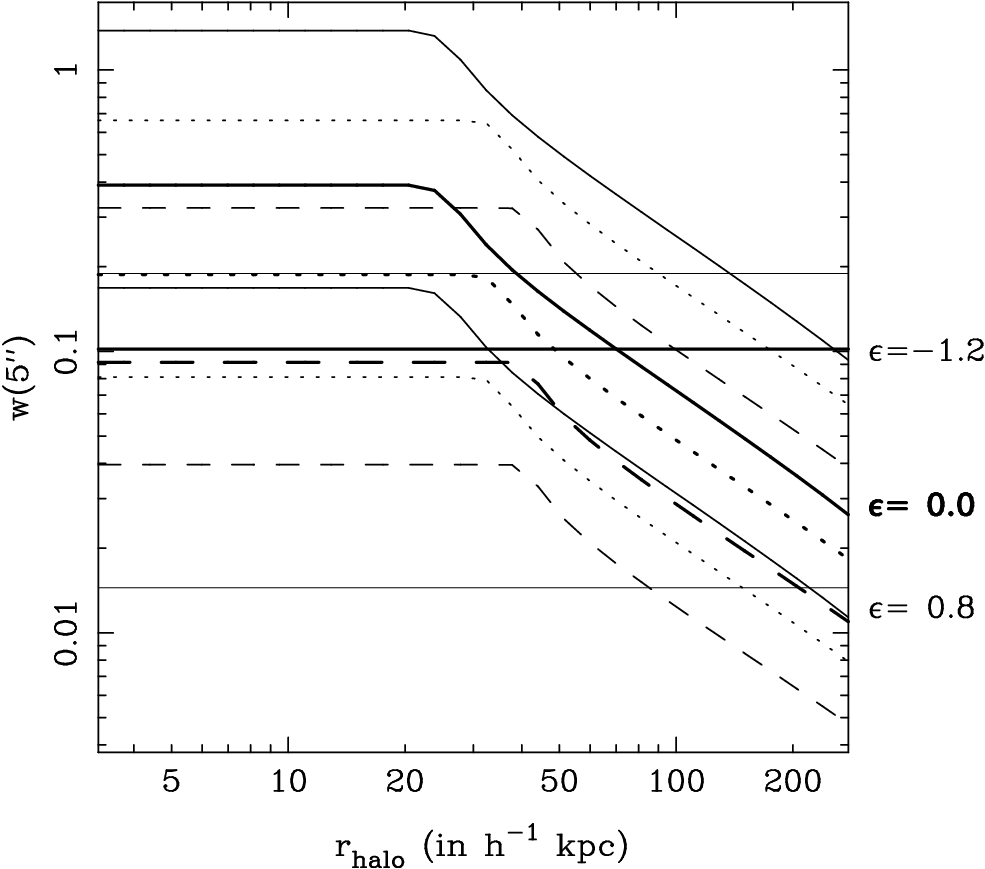"} }
}
::::
\caption{ \mycaptionfont
\label{f-wrhalo}
Angular correlation function amplitude $w(5\arcs)$ calculated using
eq.~(\protect\ref{e-limber}) and eq.~(\protect\ref{e-xieps})
for a smooth cutoff in the spatial correlation 
function $\xi$ at pair separations $\ltapprox \rhalo$, in \hkpc,
[eq.~(\protect\ref{e-xirhalo})] to take into account
the non-zero size of dark matter halo radii.
The present-day correlation length is fixed to $r_0=5\.5${\hMpc}.
Other parameters and line styles are as 
for Fig.~\protect\ref{f-wr0}. 
}
\end{figure} 
} 

\def\fwthprop{   
\begin{figure}
\centering 
\nice \centreline{\epsfxsize=8cm
\zzz{\epsfbox[51 36 460 457]{"`gunzip -c wthe0.ps.gz"} }
{\epsfbox[51 36 460 457]{"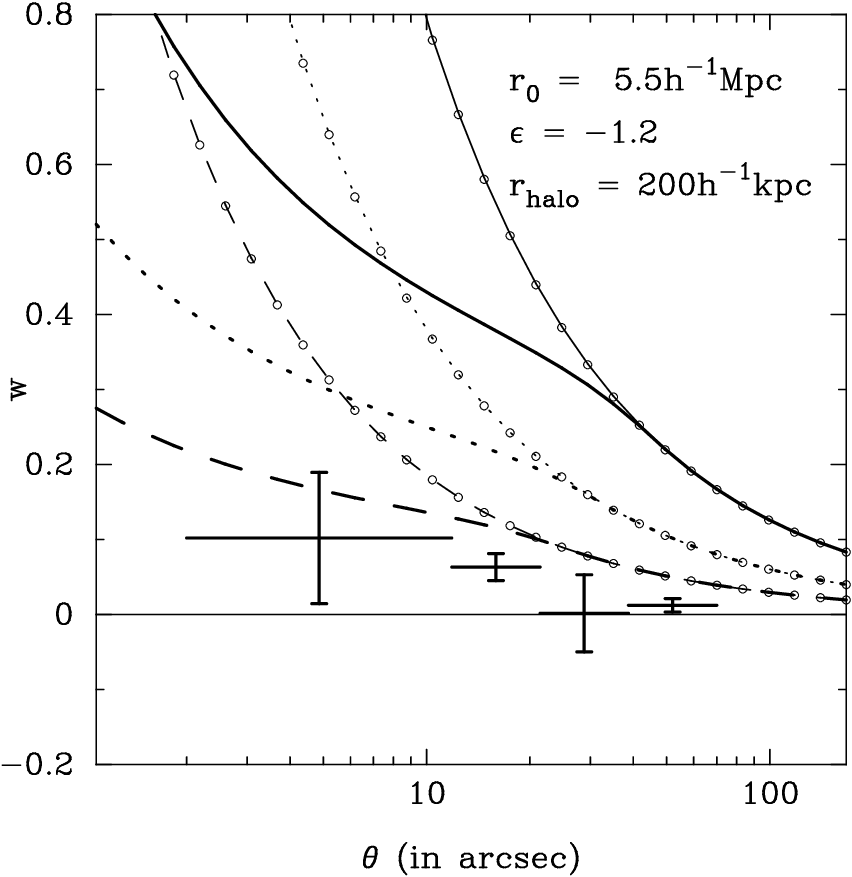"} }
}
::::
\caption{ \mycaptionfont
\label{f-wthprop}
Angular correlation functions with and without a halo cutoff
of $\rhalo=100$\hkpc, for clustering stable in proper units.
Different line styles are 
for different metric parameters as in Fig.~\protect\ref{f-wrhalo}.
Thick lines are calculated with the halo cutoff, thin lines without.
The circles are simply $1-\gamma$ power laws extrapolated from
the right-hand side of the plot.
Our HDF estimates obtained using eq.~(\protect\ref{e-wham24}) 
are also shown.
}
\end{figure}
}  

\def\fwthcomov{ 
\begin{figure}
\centering 
\nice \centreline{\epsfxsize=8cm
\zzz{\epsfbox[51 36 460 457]{"`gunzip -c wthe12.ps.gz"} }
{\epsfbox[51 36 460 457]{"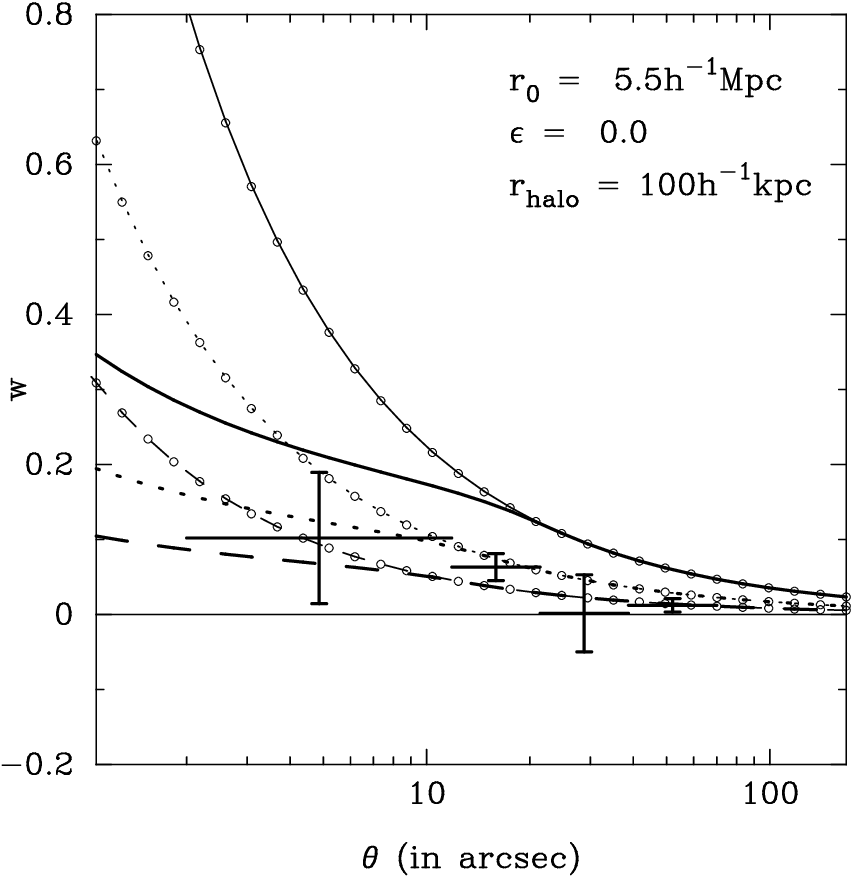"} }
}
::::
\caption{ \mycaptionfont
\label{f-wthcomov}
As for Fig.~\protect\ref{f-wthprop}, but
for clustering stable in comoving coordinates ($\epsilon=\gamma-3$) 
and for $\rhalo=200$\hkpc.  
}
\end{figure}
}  

\def\tabwamp{ 
\begin{table*}
\caption{\label{t-wamp}
Estimates of amplitude and slope of $w(z\approx2,2\arcs <\theta <40\arcs)$ 
for different techniques. Columns show number of galaxies $N_g$; 
the assumption {\bf A} for re-introducing large scale variance 
being either {\bf A}$=N$, a constraint by $\nmean$, or 
{\bf A}$=\gamma$, a constraint by assumption of a power law of slope
$1-\gamma$; 
the value of $1-\gamma$ assumed for parameters present in later
columns;
effective angle $\theta_{1-\gamma}$ (weakly dependent on assumed
slope); $w(\theta\approx 5\arcs)$ and its standard error in the 
mean $\sigma_{\overline{w}}$ estimated free from power law 
assumptions using eq.~(24) of 
\protect\citet{Ham93}; 
an upper limit to $(1-\gamma)^+$ for a power law fit to the
same derivation and its uncertainty;
$w(\theta\approx 5\arcs)$ and 1$\sigma$ uncertainties based on 
the estimator of \protect\citet{LS93};
$w(\theta\approx 5\arcs)$ and 1$\sigma$ uncertainties based on 
eq.~(15) of \protect\citet{Ham93}. 
The values of $w$ are linear, not
logarithmic.  The first two lines are for masking of large objects,
the second two without masking. 
}
$$\begin{array}{cccc ccc cccc cccc}
\hline 
\multicolumn{4}{c}{N_g} & \mbox{\bf A} 
& (1-\gamma) & \theta_{1-\gamma} &    
\multicolumn{4}{c}{\mbox{\rm Hamilton (1993) eq.~} (24)}  &   
\multicolumn{2}{c}{\mbox{\rm LS (1993)} } 			
&\multicolumn{2}{c}{\mbox{\rm Ham (1993) eq.~} (15)} \\   
\mbox{\rm tot}& \#2 & \#3 & \#4 &
&&&
w(\theta) & \sigma(\overline{w}) & (1-\gamma)^+ & \sigma[(1-\gamma)^+] &
w(\theta)_{1-\gamma} & \sigma[w(\theta)_{1-\gamma}] &
w(\theta)_{1-\gamma} & \sigma[w(\theta)_{1-\gamma}] \\
\hline
\multicolumn{14}{l}{\mbox{\rm Masking:}} \\
  142 &     55 &     35 &     52 &   N&  &  & 
   0\.10 &    0\.09 &   -0\.47 &    0\.56 \\
  142 &     55 &     35 &     52 &   \gamma& -0\.7 & 5\.5 & 
&&&&
 0\.20 &    0\.11 \\  
 142 &     55 &     35 &     52 &  \gamma& -0\.8 & 5\.3 &
   0\.16 &    0\.10 &    -0\.96 &    0\.27 & 
  0\.19 &    0\.10 &  \\
 142 &     55 &     35 &     52 & \gamma& -0\.9 & 5\.1 &
&&&&
  0\.18 &    0\.10 \\
\multicolumn{14}{l}{\mbox{\rm Without masking:}} \\
  150 &     59 &     35 &     56 &  N&   &     & 
  0\.16 &    0\.07 &  -0\.51 &    0\.44 \\ 
  150 &     59 &     35 &     56 &   \gamma&  -0\.7 &  5\.5    & 
&&&&
  0\.27 &    0\.11 \\  
 150 &     59 &     35 &     56 & \gamma&  -0\.8 & 5\.3 &
&&&&
  0\.26 &    0\.10 \\
\multicolumn{14}{l}{\mbox{\rm Without integral constraint 
correction ($C\equiv 0$):}} \\
142 &     55 &     35 &     52&  \gamma&  &&
 &&&&
  0\.09 &    0\.08 &    0\.12 &    0\.11 \\ 
\multicolumn{14}{l}{\mbox{\rm Without merging of close objects:}} \\
  153 &     61 &     35 &     57 &   N&  &  & 
  0\.12 &    0\.09 &  -0\.66 &    0\.55 \\ 
  153 &     61 &     35 &     57 & \gamma&  -0\.7 & 5\.5 & 
&&&&
  0\.19 &    0\.11 \\  
  153 &     61 &     35 &     57 & \gamma&  -0\.8     & 5\.3& 
  &&&&
   0\.18 &    0\.10 \\
\hline
\end{array} $$
\end{table*}
} 


\section{Introduction}
Structure in the Universe as represented by visible galaxies 
is commonly represented by the two-point spatial auto-correlation 
function, $\xi(r).$ This can be approximately parametrised 
as a power law 
\begin{equation}  \label{e-xi}
\xi(r) = (r_0/r)^{\gamma}
\end{equation}
where $r$ are is the spatial separation of galaxy pairs, $r$ and $r_0$
are expressed in comoving coordinates and $\gamma$ represents the
approach to homogeneity at large length scales (e.g.
\citealt{GroP77}). For a recent review of the galaxy correlation
function, see \citet{Peac97}. 

Since the major epoch of star formation in the
Universe seems to have taken place at $ z \sim 1-2$ \citep{Madau96},
this is a particularly interesting period.
It would be useful to observationally estimate $\xi$ at this
epoch in order to link gravitational theory 
with the study of luminous objects.

\citet*{Vill97} have already made an estimate of the projected, 
angular two-point auto-correlation function, $w(\theta),$ (where
$\theta$ is the angle separating two galaxies) and hence
$\xi,$ for galaxies in the Hubble Deep Field 
(hereafter, HDF; \citealt{Will96}),
to median redshifts up to \zmed$\approx 1\.9$ from an
$I$-band selected sample.  Several recent estimates for $\xi$ have
also been made for both higher redshift objects (\citealt{GiavDCP98}
at $z\sim3$; \citealt*{MPR99} in the HDF at $2\.5< z< 4\.5$) and for
lower redshift galaxies (\citealt*{Conn98} in the HDF at $z<1\.6$).

An improvement on the \citeauthor{Vill97} estimate should be possible by 
using (i) a $U$-band selected sample to remove nearly all galaxies at
redshifts $z\gtapprox 2\.5,$ and (ii) photometric redshift redshift estimation
on this sample to remove most galaxies at both lower and higher 
redshifts than $1\.5\ltapprox z \ltapprox 2\.5.$

This is the option chosen here, using \citeauthor{Mob98}'s (1998)
$U$-selected sample of HDF galaxies, dominated by ellipticals 
and starburst galaxies
at $z\sim2$ (\S\ref{s-obsdata}). 
Although galaxies selected in $U$ at $z\sim2$ are selected in rest-frame
wavelengths which do not correspond to those in which galaxies in
typical surveys are selected, the high rate of star formation at 
this epoch implies that the galaxy mix may not be too different
from other samples.

As in \citet{Conn98} and \citet{MPR99}, 
$w$ is estimated in a redshift range
limited by photometric redshifts (\S\ref{s-meth}) and Limber's equation
[equation~(\ref{e-limber})] is used to consider interpretation 
in terms of the spatial correlation function $\xi$ 
(\S\ref{s-xitheory}).

In addition, (i) the estimate of $w$ is made
without having to make a prior assumption about the shape and slope of
the correlation function (\S\ref{s-formula}) and by avoiding 
introduction of linear terms in the correction for the integral
constraint; and
(ii) the effect of bias introduced in regions of sky obscured by bright objects
 is demonstrated (\S\ref{s-mask}). 

The word `bias' in the previous sentence is used in its general
sense. Since the primary aim of this paper is observational estimation,
galaxy-to-matter bias (e.g. \citealt{Ost93,PDJ89,Matar97}) is not treated here, 
apart from discussion of a possible high redshift, high initial bias 
in \S\ref{s-dcp}.

In interpreting the $w$ estimate via integration of the relativistic version 
of Limber's equation, the size of galaxy dark matter haloes is explicitly
considered. The effective three-dimensional separations studied here, 
i.e. the median separation of galaxy pairs 
which contribute to 50\% of the projected correlation, is  
{\refffifty}$\approx25$\hkpc$-250${\hkpc} \citep{RV97}. 
Dark matter haloes must overlap at these length scales 
and it seems likely that some modification of the shape of 
the correlation function is necessary, since the dominant physics
in coexistence of galaxy pairs becomes local rather than cosmological.

The simplest correction seems to be to introduce a smooth cutoff
in $\xi$ below a characteristic halo radius. This is the approach 
adopted here (\S\ref{s-rhaloform}).

The resulting estimates of $w$ are presented in 
\S\ref{s-w}. In \S\ref{s-xi}, these are interpreted in terms of the spatial
correlation function with and without a correction for halo radii, 
for a range of simple power law hypotheses for correlation evolution
with redshift, and for a realistic 
range in values for the cosmological curvature parameters, 
$\Omega_0$ and $\lambda_0.$

Comparison with and implications from other estimates of $\xi$, 
in particular regarding the decreasing correlation period (DCP), are
provided in \S\ref{s-discuss}, and \S\ref{s-conclu} concludes.

The Hubble constant used here is $h\equiv H_0/$ $100$km~s$^{-1}$~Mpc$^{-1}.$
The spatial correlation function is presented in comoving coordinates. 
Galaxy halo sizes are discussed in proper units. The
metric parameters $\Omega_0$ and  $\lambda_0$ 
are quoted when first used (Fig.~\ref{f-wr0}).

\section{Observational Data Set} \label{s-obsdata}

The HDF is presented in detail by \citet{Will96}. 
Selection of galaxies in the 
$U$-band (strictly speaking, in the F300W band) 
for the three Wide Field Camera (WFC) fields
of the HDF is described in \citet{Mob98}, and is estimated to 
be complete to $U<27$ ($U_{AB}$ magnitudes used here). 

Photometric redshifts were calculated by \citet{Mob98} using template
spectral energy distributions (SED's) derived from
 stellar evolutionary population
synthesis models. These extend from the far-UV to 1mm, self-consistently
include stellar emission, internal extinction and re-emission by dust,
and include a range of metallicities. The synthetic SED's were
constrained both by locally observed galaxies of different morphological
types and by the HDF galaxies for which spectroscopic redshifts are
available. The r.m.s. scatter between spectroscopic and photometric 
redshifts is estimated as $0\.11.$ 

It should be noted that due to the lack of strong breaks in the optical 
passbands, the uncertainties in photometric redshifts are generally 
greater within the $1\.5 < z < 2\.5$ interval than at lower and higher 
redshifts. This is not a problem for this analysis. Since only the
projected correlation function is estimated, errors within the
interval have no effect. Scattering in the redshifts at the low
redshift end might have some effect, since although the numbers of
galaxies are relatively lower, their autocorrelation is higher, than
at high redshifts. Uncertainty at the high redshift end should
be minimised by the $U$-band selection.

\section{Method}\label{s-meth}

\subsection{Estimation of the Angular Correlation Function $w(\theta)$}
\label{s-formula}

\subsubsection{Angular limits}

Since variance on the scale of the sample can only be corrected
by information external to the sample, the estimation of $w$ at
angles similar to the size of either a single WFC field or the 
combined field cannot statistically represent structure at
that projected angular scale. The presentation of angular correlations
up to 80{\arcs} \citep{Vill97} or 220{\arcs} \citep{Conn98}
can therefore be considered as providing
investigations of individual examples of structure, which could 
be used to select samples of individual high redshift clustered
structures (or voids).

Although that approach is not without interest, the goal here
is for the study of the statistics of galaxy clustering, 
so the upper limit in angles used for correlation estimates
is $\approx 40$\arcs, half the size of a single WFC field. 
Previous experience \citep{RP94} suggests that this is a conservative
limit above which large scale variance should be excluded. For 
the $I$-band limited HDF sample of \citet{Vill97}, 
formal interpretation of the error bars of \citeauthor{Vill97}'s Fig.~1
shows that negative and positive offsets of $w(\theta_i)$ 
estimates at 25-80{\arcs} from the power law corrected fit are mostly
more significant than positive offsets on small scales. 
An alternative fit of any reasonably smooth function through
the points
would not reduce what is most simply interpreted as large scale
variance.

At small angular scales there is certainly no problem in confusion
of faint sources for $\theta \gg 0\.1$\arcs; these are Hubble Space Telescope 
data. On the contrary, the limitation is physical. 
\citet{Colley96,Colley97} have pointed out a highly significant 
excess of object pairs separated by less than about 1{\arcs} in the HDF. 
This corresponds to about 6{\hkpc} 
in separation perpendicular to the line-of-sight
(to within a factor of two depending on the
values of the metric parameters), 
in proper units. \cite{RV97} showed that if these objects were
individual galaxies with a correlation similar to or higher than that 
of equation~(\ref{e-xi}), but allowing for evolution and a wide
range of acceptable parameters, 
then the measured strength of $w$  would imply 
that their three-dimensional separations 
would typically be the same as the perpendicular 
separation, and a large majority 
of pair separations 
would be less than 15-30{\hkpc} (proper). 

\citeauthor{Colley96} also point out that 
differential bolometric surface brightness dimming
of $(1+z)^4$ for diffuse objects relative to $(1+z)^2$ for point sources 
implies that one should in fact {\em expect} high redshift optical imaging 
to split galaxies into clumps of objects such as {\HII} regions.

While a component 
of the sub-arcsecond clustering detected by \citeauthor{Colley96}
may well be real galaxy clustering [though highly biased 
as predicted by \citet*{ORY97}, apparently detected by 
\citet{SteiDCP98} and modelled by \citet{Bagla98}], the choice
adopted here is to exclude this clustering. This is done
(i) by setting a minimum angular separation of 2{\arcs} down to
which the correlation is estimated and (ii) considering objects
separated by less than $0\.5${\arcs} as single objects. 
The second of these two limits has to be significantly smaller than
the first in order to avoid introducing a bias.

In order to maximise signal to noise, large bins are adopted as 
in \cite{RP94}. The range 2-70{\arcs} is divided into six logarithmic
bins, of which the first three are regrouped into a single bin
in order to reduce the noise, and the largest angular bin is plotted
but not used in for correlation estimates or integral constraint
corrections.
The range used for correlation estimates is therefore 
2-39{\arcs}.

\subsubsection{Masking for Bright Sources} \label{s-mask}

In any image of faint galaxies, even one such as the HDF which was
designed to avoid containing any `bright' galaxies (or stars), there are 
still some galaxies which are brighter than others. It is clear
in the HDF (e.g. Fig.~8 of \citealt{Will96}) that there are some
bright galaxies which occupy non-negligible fractions of the total
solid angle. As is generally pointed out in angular correlation studies
(e.g. \citealt{Neu,RP94}), subtraction of the light from
bright objects is imperfect, both due to the light profile only being
a smooth approximation and due to the Poisson noise from the object 
adding to that of the sky. This implies that
(a) background galaxies (or foreground dwarf
galaxies) which ought to be detected after subtraction of the bright
objects may be missed (due to being subtracted away) and
(b) Poisson noise, real components of the bright object 
(e.g. a globular cluster or a knot of star formation 
if the bright object is a galaxy), cosmic rays or combinations
of the three may be mistakenly detected as 
galaxies.

Estimating the angular correlation function depends critically on the 
selection criteria being constant across the whole field (or for 
the selection function to be known precisely enough to enable a weighting
correction to be applied) and can be biased by either of these effects.
Analyses of the HDF tend to be conservative in reducing the possibility
of mistaking artifacts for galaxies, so the problem here is more likely
to be (a) than (b). 

In either case, the correction is to apply a `mask' to regions around
bright objects. In this study, use of the $U$-selected
catalogue would be insufficient since the photometric redshift 
estimates require detection in all bands and so would be affected
by bright objects in other bands. So, 
\citeauthor{Will96}'s catalogue 
is used to mask out circular areas around all objects of limiting 
isophotal areas of greater than 10~sq.arcsec. These regions
are excluded when distributing 
random points for use in equation~(\ref{e-wham15}).

\subsubsection{Different Estimators for $w$}

The estimation of the angular correlation function amplitude for small
solid angle, faint surveys normally 
suffers from the problems of small numbers of 
galaxies, of a small signal-to-noise ratio and of correcting
for variance in the number density of galaxies on the scale of
the sample observed (also known as the integral constraint). 
In this sample, the numbers of galaxies are relatively small, 
compared to 
the numbers in typical magnitude/surface brightness limited
ground-based surveys extending 
to faint magnitudes (e.g. \citealt{Ef91,Neu,PI92,Couch93,RP94,IP95} and 
\citealt{BSM95}).

However, the small angular size of the HDF implies that $w$
can be estimated at angular separations of less than 10{\arcs}, 
where it is expected to give a stronger signal, so that the
signal-to-noise ratio may be sufficiently high for a significant
estimate.

The other problem is that of
sample-scale variance which both (a) can introduce noise into
the correlation estimate and (b) needs to corrected for in order
to have a statistic which is significant in the limit of large angles.

(a) Provided that the
amplitude of this variance 
is smaller than unity, the arguments of 
\citet{Ham93} on how the noise it introduces can be reduced to a 
quadratic term rather than a linear term become valid. 
Since the three WFC fields are best considered independently 
(the small increase in the number of pairs across the borders between
does not seem to justify the possible systematic effects of that
particular sub-sample of the total close pair distribution), 
a simple, order-of-magnitude estimate of the variance is 
provided from comparison of galaxy numbers in the three fields.

\citeauthor{Ham93}'s equation (15), adapted for the angular 
correlation function, is then an estimator which minimises the
noise contribution of this variance relative to other estimators
such as that of \citet{LS93}.
\citeauthor{Ham93}'s estimator, expressed in conventional terms, is
\begin{equation}
w(\theta) = { N_{gg} N_{rr} \over N_{gr}^2 } -1
\label{e-wham15}
\end{equation}
where $N_{gg}, N_{rr}$ and $N_{gr}$ are the numbers of galaxy-galaxy
pairs, random point-random point pairs and galaxy-random point pairs
respectively, separated in each case by an angle lying in 
the interval $[\theta,\theta+\Delta \theta]$. 
\citeauthor{LS93}'s estimator, in the same terminology, is
\begin{equation}
\label{e-ls93}
w(\theta) = { N_{gg} - 2 N_{gr} + N_{rr} 
	\over N_{rr} }
\end{equation}

The random points are points selected pseudo-randomly 
from a uniform distribution in a field of the same shape and 
size as that from which the galaxies are selected. 
This corrects for edge effects.

(b) Large-scale variance is not included in 
eq.~(\ref{e-wham15}).
The common technique for re-introducing large-scale variance
is to assume that $w$ should be a power law of a given slope,
\begin{equation}
w(\theta)= A_w \theta^{1-\gamma}
\label{e-wpowerlaw}
\end{equation}
where $A_w$ is a free parameter and $\gamma$ is that of 
eq.~(\ref{e-xi}). The uncorrected estimates $\{1+w(\theta_i)\}$ 
have a constant term added (or are multiplied by a constant factor) 
such that a power law of slope $1-\gamma$ is obtained.

This assumption is that adopted by nearly all authors
of faint galaxy angular correlation function analyses who do not
have large enough fields to avoid having to make this integral 
constraint correction. \citet{Vill97} and \citet{Conn98} 
adopt this technique, and assume $\gamma=1\.8.$ 

Some authors only make the assumption that $w$ 
is a power law, and try to estimate both the amplitude 
and slope of the power law by $\chi^2$ minimisation 
(e.g. \citealt{Neu}).

This assumption is obviously sensitive to any minor changes in 
the slope or the shape of the correlation function over this 
limited angular range. 
As reviewed by \citet{Peac97}, it is
likely that $\xi$ is not perfectly fit by a single power law.
Either a double power law or a fitting function such as that
of \citet{Ham91} are likely to be closer to the intrinsic
shape of $\xi.$ The effective slope over a small range in angle
may be different enough from the assumed value of $\gamma$ to
add significant systematic error to the amplitude estimate.

Moreover, this corrected formula reintroduces linear error terms
into the estimate. Effectively, the estimator becomes
\begin{equation}
\label{e-ls93ic}
w(\theta) = { N_{gg} - 2 N_{gr} + N_{rr} 
	\over N_{rr} } + C
\end{equation}
for some constant $C.$

\citeauthor{Ham93} presents a version of this equation, [his eq.~(24)],
which optimally corrects for the large-scale variance:
\begin{equation}
\label{e-wham24}
w(\theta) = { N_{gg} - 2 \nmean N_{gr} + \nmean^2 N_{rr} 
	\over \nmean^2 N_{rr} }
\end{equation}
where $\nmean$ is 
the mean number density 
estimated by some means external to the sample, divided by the
number density of the sample itself.

It is clear that varying the value of $C$ 
in eq.~(\ref{e-ls93ic}) is not equivalent to varying 
$\nmean$ in eq.~(\ref{e-wham24}). Finding a best fit for 
$w(\theta)$, subject to the constraint that it has a certain
shape and slope, will not give the same results in the two equations.

In addition, 
\citeauthor{Ham93} suggests using large scale variance estimated
from very large scales such as the cosmic microwave background 
measurements by COBE as the external constraint. 
However, the extrapolation over time and 
space from COBE measurements to the present work would obviously 
require numerous assumptions. 

Since in the present case the three WFC fields are analysed
independently, a simple way of estimating the $\nmean$ in the present
case is to use the mean density of the three fields as the value of
$\nmean$ to use in evaluating eq.~(\ref{e-wham24}) for any single field.
This should correct for variance on a scale considerably larger
than that of interest. 

As is seen below, this results in a mean
correlation function which is positive and consistent with 
a power law, {\em and additionally avoids having to assume that 
it should be a power law in order to make the correction}.

In principle, there should still be a bias from variance on yet
larger scales, but without large-scale galaxy surveys at $z\sim2$ 
there is no obvious way to make this correction directly from
observational data, apart from adopting the conventional practice
of assuming that $w(\theta)$ should be a power law of a given slope.

The uncertainty in the correlation estimates is itself estimated, for
any angular bin, by the dispersion in the $w$ estimates between the
three WFC fields.

\fwimage   

\fwimagenohol 

\subsection{Relating $\xi$ to $w$} \label{s-xitheory}

\subsubsection{Use of Redshifts in Limber's Equation}\label{s-lim}
The angular correlation function (small angle approximation) 
is given by the the double integration of $\xi(r,z),$ 
\begin{equation} w(\theta,N_z) = 
{ \int dz\, N_z(z)^2 \int du\, \xi(r,z)
	\over
 \left[\, \int dz\, N_z(z) \,\right]^2}
\label{e-limber}
\end{equation}
where $z=(z_1+z_2)/2$ and $u=z_1-z_2$ parametrise the redshifts of two galaxies
at redshifts $z_1$ and $z_2,$ $r(z,u)$ is the spatial separation of
the two galaxies, and 
$N_{z}(z)= $d$N/$d$z$ is the redshift 
distribution of the sample studied \citep{Lim53,Ph78,Bible,Ef91}.

In the present case, \citeauthor{Conn98}'s (1998) method of modelling
a redshift distribution by summing the individual photometric redshifts,
represented as Gaussian distributions centred on the estimated values,
where $\sigma_z=0\.11,$ is adopted.

\subsubsection{Low Redshift Contribution to Limber's Equation} 
\label{s-lowzlimber}
However, as remarked upon by \citet{RV97}, the physical scales of 
the function $\xi$ which contribute to the angular signal need to
be carefully considered. 

In principle,
the combination of very short length scales and very high correlations
in the very low $z$ component of the cone of observation 
can be a problem in the 
integration of Limber's equation (whether analytical or numerical).  
This is because although the volume at low redshifts 
is small, the correlation modelled by a power law at small separations
becomes extremely high\footnote{It was pointed out by Bruno Maillard 
(personal communication) that since the observer lives in a 
non-random point of space, i.e. in a group and local super-cluster,
the correct modelling of the correlation should be even higher towards
zero pair separations at near zero redshifts.}. The resulting contribution
to $w$ can be non-negligible (see Fig.~1, \citeauthor{RV97}).

Although this could be considered to have 
some physical meaning, e.g. to represent the clustering of very faint
($U\ltapprox 27$) dwarfs in the Local Group, the small angular size
of the field ($\approx 4\.5\e{-7}$~sr) implies that any celestial population
of objects needs to have at least $\approx 2\.8\e{7}$ objects over
$4\pi$~sr in order to expect (in the mean) a single object in the HDF.
So a few small dwarf spheroidals distributed on the scale of the local
supercluster could possibly be present in the HDF, but the presence
of a Local Group galaxy would only be due to a very rare event.

In the present case, the interest is in cosmological distances, 
so photometric redshift estimation excludes this very low redshift
component --- certainly 
in the calculation and most probably in the observational data. 

\subsubsection{Galaxy Halo Sizes}  \label{s-rhaloform}
Given the exclusion of low redshifts ($z\ltapprox 1\.5$),
the minimum three-dimensional separation considered in the integral 
is not too different from the perpendicular separation at the median
redshift of the redshift band $1\.5 \ltapprox z \ltapprox 2\.5.$ 
The perpendicular separations over these redshifts are considerably 
smaller than 100-200\hkpc, i.e. than likely present-day sizes of typical
$L^*$ galaxy 
haloes\footnote{The word `haloes' is used here in the galaxy formation
sense of dark matter haloes containing baryonic and/or nonbaryonic 
non-luminous matter.}. 

Is it physically reasonable 
for galaxy pairs to exist as close to one another as 6{\hkpc} for 
long enough to have a chance of being observed in an HDF-like sample?

Hierarchical galaxy formation models would imply that galaxy
haloes should be smaller at $z\sim2$ than at the present.
However, at typical redshifts of Lyman-$\alpha$ and
metal line absorbers in front of quasars, 
the haloes of galaxies are estimated as having gaseous radii 
of around 50-200{\hkpc} 
\citep{BeBo91,Bech94,Lanz95,FDCB96,LeBrun96,Chen98}.

It is dynamically unlikely that the dark matter halo radii
could be smaller than the gaseous radii, so the closest fit
between theory and observation would be for dark matter halo
radii to be about the same size as the gaseous radii.

It could be possible that the $z\sim2$ galaxies in the HDF are mostly
dwarfs (defined by mass), brightened by bursts of star formation,
and that the types of galaxies corresponding 
to quasar absorbers are very rare in the HDF. In this case, galaxies 
could
indeed coexist according to an extrapolation of $\xi$ to small pair
separations. Alternatively, since the half-light radii of HDF galaxies
are small, it could be possible that the baryonic, stellar cores
of haloes, i.e. galaxies, are tightly enough bound that they co-orbit
in the haloes (i) without merging over a Hubble time and 
(ii) they do this is such a way that their pair separations 
remain statistically well modelled by $\xi.$

However, it seems reasonable that a fair fraction of the HDF
galaxies, which are detected in spite of $(1+z)^4$ bolometric 
surface brightness
dimming, have halo radii (gaseous and dark matter) 
in the 50-200{\hkpc} range. 

Once gaseous haloes overlap, it is likely that in some fraction of cases
(e.g. low relative velocities), these haloes will merge in much
less than a Hubble time. Among the halo pairs that merge, some 
fraction of their baryonic, `galaxy' cores will also merge rapidly.
This removes pairs from the pair distribution. So, the pair 
probabilities, hence $\xi$, should be lower than that expected from 
the low $r$ extrapolation of $\xi$ from its cosmological context.

It is interesting to note the argument of \citet{BB96} that 
dynamical estimates (from the cosmic virial `theorem') are consistent
with dark matter halo radii larger than conventionally assumed, 
i.e. bounded below by 300{\hkpc}. However, there is no indication 
that matter extending out 
to these large radii would be baryonic, i.e. collisional,
so rapid merging of pairs would be unlikely. In such a scenario,
the expression `halo radius' could be replaced by 
`halo impact parameter leading to rapid merging'.

Many possible effects of pair exclusion 
are probably convolved, but the 
observational constraints are weak. So, the correction
adopted here is simply to include a cut-off in $\xi$ by a single parameter
representing a `typical halo radius', $\rhalo.$ We consider how
this affects the interpretation of the observed values of $w.$
The cutoff is modelled smoothly as
\begin{equation}
\xi' \equiv \xi\;\ttimes \left\{ \begin{array}{lll}  1 , & r/(1+z) \geq \rhalo \\
	\exp \left\{ { - {[\rhalo - r/(1+z)]^2 
	   \over 2 \sigma^2}  \mbox{\rule{0ex}{2ex}} } \right\}, 
						& r/(1+z) < \rhalo 
		\end{array} \right.
\label{e-xirhalo}
\end{equation}
where $\rhalo$ is in {\em proper} units, $r/(1+z)$ is the 
proper separation corresponding to $r$ in comoving
units, 
and $\sigma=\rhalo/2$. This is a Gaussian cutoff,  
so that the probability of a pair of galaxies 
existing at $r/(1+z)= \rhalo/2$ is multiplied by $e^{-1/2} \approx 0\.61,$ 
and that of a pair 
at $r/(1+z)=\rhalo/10$ is multiplied by $e^{(-9/5)^2/2} \approx 0\.20.$

\fwiformtwo  

\fwiformtwonohol

\fwiformtwonomg

\fwiformtwofit 


\fwiformthrB

\fwiformone

\tabwamp

\section{Results} \label{s-res}
Galaxies in \citeauthor{Mob98}'s (1998) HDF catalogue which have 
redshifts estimated as $1\.5 \le z \le 2\.5$ 
are shown in Fig.~\ref{f-wimage}, 
for apparent magnitudes
$23 \le \UFthhW \le 27$, over which the catalogue should be
complete. Galaxies in areas biased by large objects are excluded.
The galaxy positions are those of 
\citeauthor{Mob98}, the galaxy images are from the 
the HDF archival F300W image and 
\citeauthor{Will96}'s (1996) catalogue is used 
for the positions of large
objects. 

To be conservative, objects detected at less than 
\begin{equation}
\thmask \equiv 2 \sqrt{A/\pi},
\end{equation}
from the centre of a large object, 
where $A=10$~sq.arcsec is the solid angle within the isophotal limit of the large
object, are excluded. \citeauthor{Will96}'s catalogue includes
several estimates of the isophotal areas for the brightest objects,
depending on how many components of what seems to be a single object
are counted as a single one. This is seen asseveral concentric rings 
around a single bright object. The masked zone in such a case 
is defined by the largest ring.

Fig.~\ref{f-wimage_nhol} shows a similar image, but without any
masked regions.
It is clear by comparison of the two images that several of the 
`voids' in this figure correspond to the larger of the masked
regions in Fig.~\ref{f-wimage}. This suggests that galaxies have
been missed in these regions. 

\subsection{Estimates of $w$} \label{s-w}

Fig.~\ref{f-wiform2} shows the angular correlation function estimated
using eq.~(\ref{e-wham24}) for the selected galaxies shown in
Fig.~\ref{f-wimage}. Large-scale variance has been introduced via an
estimate of $\nmean$ `external' to the sample, as recommended by 
\citet{Ham93}, without making any assumption about the shape or 
slope of $w$ 
This is done 
by considering the three WFC fields to be independent samples,
and $\nmean$ to be the global mean density for the
three fields. 

Strictly speaking, large scale variance at scales larger than
the combined WFC fields would still be missing from our 
estimate of $\nmean.$
The standard deviation
of the number of objects per WFC field is $10\.8$ 
(with masking and with merging of close objects),  
and the mean is $47\.3,$ i.e. the standard error in the mean
of the $\nmean$ estimate 
is about 13\%. 

One way of potentially reintroducing the large scale 
variance is to follow the convention of fitting to a power law of a given
shape and slope. This would only be correct if the true shape and slope 
are in fact those chosen.
This is discussed further in 
\S\ref{s-convest}.

The three fields, considered as independent samples, 
are averaged and the point 
in the lowest angular bin (at $\theta\approx5$\arcs) is considered
to be the estimate of the amplitude of $w.$ A fit to the three points
in $2\arcs < \theta < 39${\arcs} is considered as an upper limit
to the slope $1-\gamma.$

Table~\ref{t-wamp} lists numerical values for this estimate and 
for other estimates which take into account 
possible systematic errors and conventional estimators for $w.$
The median angle in the $\theta\approx5${\arcs} bin is also
listed. Since this depends on the slope $1-\gamma$ and is 
only slightly higher than 5{\arcs}, the bin can be referred
to as the `5{\arcs} bin' for brevity.

\subsubsection{The Effect of Masking}

Comparison of Fig.~\ref{f-wiform2} and Fig.~\ref{f-wiform2nh} 
shows that several `voids' in the unmasked galaxy
distribution shown in Fig.~\ref{f-wimage_nhol} correspond to masked areas 
shown in  Fig.~\ref{f-wimage}. This has a noticeable 
effect on $w$: the value increases
by about 60\% if the `voids' are considered to be unbiased regions.  
The masked and unmasked values of $w(5\arcs)$ are 
$0\.10\pm0\.09$ and $0\.16\pm0\.07$ respectively 
(Table~\ref{t-wamp}).

Is the difference statistically significant? Suppose we consider
the masked value, $0\.10,$ to be the true value, the results of 
repeated experiments to be normally distributed about this 
with a standard deviation of $0\.09$, and the unmasked value,  
$0\.16,$ to be a realisation of the same experiment. 
Then the null hypothesis that the unmasked value comes from
the same experiment can only be rejected at a
$\Pgauss[(|w- 0\.10|/0\.09)\sigma < (0\.06/0\.09) \sigma] = 50\%$ 
confidence level. So it is not proven to high significance 
that objects have indeed been `hidden' by the large objects.

The difference that does exist can be understood arithmetically
as follows.
An uncorrelated distribution punctuated by voids would become 
correlated; an already correlated distribution generally 
becomes more so.
So, the effect is strongest for the two bins smaller than the typical 
`void' size, as expected.

\subsubsection{The Effect of Object Multiplicity}

The effect of object splitting is small. Fig.~\ref{f-wiform2nm} 
shows $w$ estimated as in Fig.~\ref{f-wiform2}, but by considering
objects separated by less than $0\.5${\arcs} as genuinely 
independent galaxies. If a single galaxy is seen as, say, 
multiple {\HII} regions, which are wrongly assumed to be independent
galaxies, than the correlation 
function would in effect be weighted more strongly by (a) galaxies 
close to these `multiple galaxies' and (b) pairs of such `multiple
galaxies'. 

Effect (a), on small scales, would be noticeable if the multiple galaxies are
located in clustered regions, in which case they pair up with close
galaxies. The multiplicity may be due to either the `halo building
blocks' or the `{\HII} region' interpretations.

If the multiplicity is due to the `halo building blocks'
model, then the `multiple galaxies' would be in clustered
regions, since there is no sharp cut in the hierarchy of 
clustering. In this case, effect (a) should be expected. 

On the other hand, in the `{\HII} region' interpretation, 
due to $(1+z)^4$ bolometric surface brightness dimming, {\HII}
regions are favoured over disks relative to shot noise.
In this case, there 
is no reason to expect these galaxies to be in clustered regions
more often than `non-multiple galaxies' are in clustered regions,
apart from the extent to which star formation requires close
galaxy-galaxy interactions and/or merging. 

So the presence or absence of this effect would respectively favour the 
`halo building block' or the `{\HII} region' interpretations
of the sub-arcsecond pair excess noticed 
by \citet{Colley96}.

Effect (b) would be stronger as long as at least two
`multiple galaxies' are in the sample, since it would depend on
the square of the r.m.s. multiplicity. If the multiple 
objects are relatively rare, the effect
would occur at relatively large separations. The effect would be
to increase the correlation at the separations of the multiple objects.
If the multiple objects were common enough, then their contribution
would reflect their intrinsic correlation as a population.

In this case, the multiple objects are rare: 
6, 0 and 5 objects are doubly counted in the three fields respectively 
if objects separated by less than $0\.5${\arcs} are considered as
independent real galaxies (see Table~\ref{t-wamp}).

Hence, effect (b) is to be expected only at large scales and is separable
from effect (a). 

Between Figs~\ref{f-wiform2} and \ref{f-wiform2nm}, 
the effect is small in the 5{\arcs} bin and most obvious in the 
21-39{\arcm} bin. This favours the `{\HII}' region interpretation, 
though not significantly.

\subsubsection{Conventional Estimators} \label{s-convest}

\citet{Vill97} and \citet{Conn98} do not use
an estimate of the mean density $\nmean$ in order to include 
large-scale variance.  They adopt the conventional technique 
of first estimating $w$ with the mean density of the sample,
then searching for an additive correction which best transforms the set
of binned $\{w_i\}$ values close to a power law of slope
$1-\gamma=-0\.8.$ 

In the case of eq.~(\ref{e-wham24}), the only means of modifying this
to fit a power law is to recalculate $w$ for different values of 
$\nmean$ until the closest resulting set $\{w_i\}$ to a power law
of desired slope is obtained.
The result obtained with the fixed `external estimate'
of $\nmean$ gives a slightly shallower upper limit
slope,  $1-\gamma=-0\.47\pm0\.56,$ than the conventional 
$1-\gamma=-0\.8.$ So, 
it could be expected that this more conventional
constraint would give a slightly lower value for $w(5\arcs),$ 
in order to increase the steepness of the slope.

However, this is not the case here; eq.~(\ref{e-wham24}) is not dependent
on a simple additive parameter as is the case of eq.~(\ref{e-ls93ic}).
Fig.~\ref{f-wiform2fit} and the entry in Table~\ref{t-wamp} show
that the closest slope to $1-\gamma=0\.8$ 
attainable is $1-\gamma=-0\.96,$ and that 
$w(5\arcs)$ increases from $0\.10$ to $0\.16.$

The formal significance of the difference obtained by 
using the power law constraint 
is (by coincidence) 
numerically the same as that if masking is omitted. 
That is, if the fixed constraint determines the true value
of $w(5\arcs),$ then the $1-\gamma=-0\.8$ power law fit gives a
value of $w(5\arcs)$ rejected at only a 
$\Pgauss[(|w- 0\.10|/0\.09)\sigma < (0\.06/0\.09) \sigma] =50\%$ 
confidence level.

This is not equivalent to using eq.~(\ref{e-ls93ic}) with the
same power law, 
as can be seen from the equations, from 
Fig.~\ref{f-wiform3b} and from Table~\ref{t-wamp}. 
For the present data, eq.~(\ref{e-ls93ic})
overestimates $w.$ 

Again, consider $0\.10\pm0\.09$ to represent the 
true value and true error distribution, and consider the estimate
from eq.~(\ref{e-ls93ic}) to be an unbiased realisation of 
this distribution. Then this null hypothesis is only rejected 
at a
$\Pgauss[(|w- 0\.10|/0\.09)\sigma < (0\.09/0\.09) \sigma] =68\%$ 
confidence level.

However, 
as mentioned above, not all large scale variance is restored by
using the mean of the three WFC fields to estimate $\nmean.$
So, an alternative is to consider the power law corrected estimate
using eq.~(\ref{e-wham24}) to be the true estimate, restoring
the large scale variance. In that case, the (power law corrected) 
estimate from eq.~(\ref{e-ls93ic}) considered as a realisation 
of the same experiment can be rejected only at a
$\Pgauss[(|w- 0\.16|/0\.10)\sigma < (0\.03/0\.10) \sigma] =24\%$ 
confidence level.

This is not at all a significant difference, which is not
surprising. The differences between the two formulae are in 
the linear error terms, so should be of the same order of
magnitude as the uncertainties.

An advantage of eq.~(\ref{e-ls93ic}) is that for a typical 
data set, uncorrected correlations [$C=0$ in eq.~(\ref{e-ls93ic})] 
are close to zero at large angles. This gives a good chance of 
being able to find a best fit power law as close as possible to
the desired slope, since $C=0$ gives a very steep slope, $C\gg1$ 
gives a very shallow slope, and the effect should be smooth and
continuous. 

This is equally a disadvantage, since the desired slope is 
attained even though the equivalent calculation, 
avoiding linear
terms, using eq.~(\ref{e-wham24}), shows that this should not be
the case, apart from consideration of the uncertainties. 
In other words, the desired slope is attained through inclusion of
linear error terms, which is obviously less than optimal.

So that readers can compare with other works, the results of using
eq.~(\ref{e-ls93ic}) for various assumed slopes and including the
effects of not masking, of setting $C=0$ and of not merging close
objects are also listed in Table~\ref{t-wamp}.

For completeness, the application of \citeauthor{Ham93}'s eq.~(15) 
[eq.~(\ref{e-wham15}) here], which avoids large scale variance 
altogether, is shown in Fig.~\ref{f-wiform1}. The 
missing variance is clear at the larger angles here.

\fzdis

\fwrzero

\subsection{Interpretations in Terms of $\xi$} \label{s-xi}

\subsubsection{Redshift Dependence}

Although the purpose of this paper is less ambitious than the study of 
evolution of the correlation function, eq.~(\ref{e-xi}), the redshift
range under study is large enough that a simple power law 
model [in $(1+z)$] of 
spatial correlation evolution needs to be discussed. 
The following evolutionary 
version of eq.~(\ref{e-xi}) is therefore adopted:
\begin{equation}
  \label{e-xieps}
\xi(r) = (r_0/r)^{\gamma} (1+z)^{-(3+\epsilon -\gamma)}
\end{equation}
\citep{GroP77}, with $r$ and $r_0$ in comoving coordinates as before,
and $\epsilon$ a factor representing evolution.

As argued by \citet{Peac97}, a double power law may provide a more
observationally and theoretically justified fit, or as shown by 
\citet{Ham91}, a rational polynomial fitting function based on 
dark matter only $N$-body simulations can provide a theoretically 
justified model. 

Given the uncertainties in the HDF data, it seems prudent to 
simply adopt the power law model. Moreover, since the clustering
is at a strongly non-linear scale, it should be expected that 
it is stable in proper units, in which case eq.~(\ref{e-xieps})
with $\epsilon=0\.0$ should be a fair approximation.
This is the possibility discussed primarily here, though other
possibilities such as the inclusion of the effects of bias could, 
in principle, also be considered.

Other possibilities for correlation function evolution include
$\epsilon=\gamma-3$ (stable clustering in comoving coordinates),
$\epsilon=0\.8$ (linear growth for $\Omega_0=1, \lambda_0=0$) 
and higher values for $\epsilon$ in the transition zone between
linear and non-linear regimes.

The smoothed redshift distribution of \citeauthor{Mob98}'s catalogue
is shown in Fig.~\ref{f-zdis}. For the differences in the correlations
between the three WFC fields to be attributed to redshift evolution,
this would have to be in the sense of stronger correlations at
higher redshifts \citep{ORY97}. Noise within the error bars shown
would be a more conservative interpretation.

\subsubsection{Zero Redshift Correlation Lengths $r_0$}
\label{s-r0res}

The correlations at redshift zero for proper length stable 
clustering (and for two other values of $\epsilon$) are shown in 
Fig.~\ref{f-wr0} for $1-\gamma=-0\.8.$ As is well known 
\citepf{YPT93}, lower values of $w$ are expected for low density 
and for low density $\Lambda$-dominated metrics than for 
an $\Omega_0=1, \lambda_0=0$ universe. 

The $w(5\arcs)$ estimate for this `no-evolution' model 
is equivalent to $r_0 = 2\.6^{+1.1}_{-1.7}${\hMpc} ($\Omega_0=1, \lambda_0=0$),
$r_0 = 3\.9^{+1.6}_{-2.6}${\hMpc} ($\Omega_0=0\.1, \lambda_0=0$) or
$r_0 = 5\.8^{+2.4}_{-3.9}${\hMpc} ($\Omega_0=0\.1, \lambda_0=0\.9$).

The high density metric implies about 2$\sigma$ inconsistency 
with low redshift estimates of $r_0 \sim 5\.5${\hMpc}
(e.g. \citealt{DavP83,Love92}), but the low density metrics are clearly
consistent with proper length stable clustering.

\fwrhalo

\fwthprop

\fwthcomov

\subsubsection{Halo Radius Cutoff} \label{s-rhalocut}

If the non-zero size of halo radii is taken into account, using
eq.~(\ref{e-xirhalo}) and $r_0=5\.5${\hMpc} 
the measured value of $w(5\arcs)$ implies halo characteristic 
radii which are very reasonable. This can be seen in 
Fig.~\ref{f-wrhalo}

For clustering stable in proper units and this 
fixed value of $r_0$, halo radii are
$\rhalo = 70^{+420}_{-30}${\hkpc} ($\Omega_0=1, \lambda_0=0$),
$\rhalo < 350${\hkpc} (1$\sigma$ upper limit; 
$\Omega_0=0\.1, \lambda_0=0$) and
$\rhalo < 210${\hkpc} (1$\sigma$ upper limit; 
$\Omega_0=0\.1, \lambda_0=0\.9$). 
Because 
no pairs of galaxies can be seen at less than the separation
perpendicular to the line of sight, $\rhalo$ has no effect below a
certain value, so for low density metrics only upper bounds are
found.

These numbers are totally consistent with estimates from quasar
absorption systems, if these are samples of a similar population 
of objects, of which some fraction merge quickly once the 
gaseous haloes overlap. 

If the $w(5\arcs)$ value estimated from the HDF
had been much lower, say, a factor of ten lower, then the halo
radii would have been considerably higher. By extrapolation, they
would have been 
$\rhalo= 680^{+3700}_{-310}${\hkpc},
$\rhalo= 500^{+3160}_{-240}${\hkpc} and
$\rhalo= 300^{+2010}_{-140}${\hkpc} respectively 
for the three metrics, suggesting larger galaxy exclusion radii 
than the estimated sizes of the absorption systems.

\subsubsection{Angular Dependence of $w(\theta)$} \label{s-angdep}

The angular dependence of $w$ might be thought to have a low $\theta$
cutoff in the presence of a non-zero value of $\rhalo.$ 

Fig.~\ref{f-wthprop} shows that this is not the case. 
A simple round number of $\rhalo=100${\hkpc} was chosen for 
illustration, for the three choices of metric and no clustering evolution
in proper units.
This cutoff clearly has an effect. However, because $w$ is a projection 
of $\xi,$ pairs of galaxies separated by $r/(1+z) > \rhalo$ 
and whose separation vectors are oriented towards the line-of-sight
are seen with perpendicular separations less than $\rhalo,$ 
so there is still a strong signal. 

This figure also shows that the high density metric model 
provides a less good fit than expected solely from considering
$w(5\arcs)$ as an amplitude estimate. An $\rhalo=100${\hkpc} 
halo size brings it into nearly 1$\sigma$ consistency with 
the $w(5\arcs)$ estimate, but it 
is inconsistent with the \mbox{12-21{\arcs}} bin.

Of the three metrics used for the figure,
the low density metrics provide the best fits. Since 
perpendicular distances for a given redshift and angle are smaller
for these metrics than for a high density one, the effect of $\rhalo$ 
becomes negligible.

Alternative ways to best fit all points would be to 
increase $\epsilon$ to a value higher than 
that for stable clustering in
proper units, i.e. $\epsilon > 0$ (e.g. \citealt{Ef91,BSM95}),
or to consider a bias factor (e.g. \citealt{Ost93}). 

However, the uncertainties shown in the amplitudes of $w(\theta)$ 
in the figure are only statistical uncertainties. 
The differences in Figs~\ref{f-wiform2}-\ref{f-wiform1} suggest 
that systematic uncertainties in the second and third bins could be too
large for detailed comparison of different metrics or models of bias.

How strong could the effect of $\rhalo$ could be? 
For stronger evolution with redshift
than $\epsilon=0,$ the expected values of $w$ would be lower
than in Fig.~\ref{f-wthprop}, so non-zero $\rhalo$ would either
have a statistically insignificant effect or would imply lower
values than observed.

A lower limit to evolution is that of
clustering which is constant in comoving coordinates.
Fig.~\ref{f-wthcomov} shows this case, for a round value 
of $\rhalo=200${\hkpc}. This is sufficient to bring 
a high density metric model to within $\sim 3\sigma$ of the
$w(5\arcs)$ value, but it remains highly inconsistent with
$w$ at all larger angles. The $\Lambda$-dominated metric model
is brought within the 1$\sigma$ limit of the
$w(5\arcs)$ value, but still remains inconsistent at about 2$\sigma$ with
the \mbox{12-21{\arcs}} bin.

\section{Discussion} \label{s-discuss}

How do these results compare with the other HDF analyses?

\subsection{Comparison with \protect\citet{Vill97}}

Table~1 of \citet{Vill97} shows that their estimate of $w$ which could
be most closely compared to the present one would be for 
their $R<29$ magnitude limited sample, which is modelled to have
a median redshift of {\zmed}$\approx 1\.9,$ similar to that of
our $U$-band plus photometric redshift selected sample. 
Since the redshift distribution of \citeauthor{Vill97}'s 
sample is much wider, a lower amplitude should be expected.

Their amplitude is equivalent to $w(5\.3\arcs) = 0\.017\pm 0\.009$
(using the effective angle over our `5{\arcs} bin' for $1-\gamma=0\.8$).
Because of the way that angular diameter changes with redshift, 
the three-dimensional separations probed by \citeauthor{Vill97}'s 
sample should not be very different from that of our sample. 
So, these separations should
be mostly in the strongly non-linear regime, where no evolution in
proper units is expected ($\epsilon=0$).

Therefore, most of the difference in amplitude should come from the
effect of the superimposition 
of many `independent' slices, each intrinsically auto-correlated 
($\xi_{ii}=\xi$)
but having zero correlation with other slices ($\xi_{ij}=0$).
An order of magnitude estimate of this effect would be to take the
half-maximum points of the redshift distribution in Fig.~\ref{f-zdis}
as the limits of $dN/dz$ used here, i.e. $1\.6 \ltapprox z \ltapprox 2\.2$, 
to take {\zmed}$/2$ and $3${\zmed}$/2$ as the limits of 
\citeauthor{Vill97}'s distribution for $R<29,$ and count the number 
of `slices' of our sample that would fit in theirs (using comoving
distance coordinates). The number of our `slices' which would fit
in theirs are $N=3\.4$-$3\.2$ (depending on the metric parameters in the
order listed above). Their correlation amplitude can then be
summed over pairs of slices ${ij}$ as 
\begin{equation}
\label{e-crude}
\begin{array}{lll}
& \sim &{
N \xi_{ii} + N(N-1)\xi_{ij} \over  N^2 \xi_{ii}} 
=  {N\xi_{ii} \over N^2\xi_{ii}} 
= {1\over N} 
\end{array}
\end{equation}
times our value, i.e. around $1/3\.3$ times ours.

In fact, their $w$ value is about $1/7$ of ours, so would be equivalent to
roughly half of ours if restricted to a single `slice'. 
For $\gamma=1\.8,$ this would imply that their inferred value of
$r_0$ (for a given set of metric parameters and no proper coordinate
evolution) would be $\approx 2^{-1.8}\approx 0\.3$ times our value.
For $\Omega_0=1,\lambda_0=0,$ our estimate of 
$r_0 = 2\.6^{+1.1}_{-1.7}${\hMpc} 
would become $r_0= 0\.75^{+0.3}_{-0.5}${\hMpc}.

\citeauthor{Vill97}'s 
Fig.~3 (left panel) suggests that 
their modelling via Limber's equation 
would imply about $r_0\sim 1\.8\pm0.5${\hMpc} for $\Omega_0=1,\lambda_0$,
$\epsilon=0,$ so this rough estimate from eq.~(\ref{e-crude}) underestimates
the integrated value by less than half an order of magnitude.

The ratio between \citeauthor{Vill97}'s $r_0$ value and ours
is about a factor of $0\.7.$ Although this disagreement is only 
significant at $\sim 1\sigma,$ 
and the two data sets are selected quite differently, 
both data sets do consist of HDF galaxies, so it 
is worth commenting on possibilities for a systematic rather
than random error.

\citeauthor{Vill97} have 1559 galaxies; we have 142. 
\citeauthor{Vill97} clearly have an advantage in numbers which should
reduce Poisson error. 

They also have the property of selecting galaxies
in rest-frame wavelengths much closer to those of major low redshift
galaxy surveys than of our $U$-selected sample. This would be an advantage
if $z\sim2$ galaxies were statistically similar to low redshift galaxies,
but less so if a large proportion of $z\sim2$ galaxies are either 
starbursting galaxies or ellipticals with UV upturns.

\citeauthor{Vill97} also have the disadvantage of combining galaxies
of widely differing redshifts, which should increase noise, 
and of not knowing the precise shape 
of the redshift distribution. This could very easily provide a 
systematic error of a factor of $0\.7$ in $r_0$, 
e.g. see Fig.~2 [panel (a)] and Fig.~3 of \citet{RV97}.

Other possible systematic uncertainties 
are the inclusion of linear uncertainty terms 
and the correction for regions obscured by large objects.
\citeauthor{Vill97} used  eq.~(\ref{e-ls93ic}) rather than
eq.~(\ref{e-wham24}) and adopted the power law of slope $1-\gamma=0\.8$
constraint. They did not refer to masking for bright objects.

\subsection{Comparison with `Low' Redshift Samples}

\citeauthor{Conn98}'s (1998) $w$ estimates are based on photometric redshifts
of 926 $\IFeofW < 27 $ galaxies in the HDF and $0\.4 < z < 1\.6.$
For $\epsilon=0,$
over the range $0\.2< \Omega_0 < 1, \lambda_0 \equiv 0,$ 
the authors infer $r_0 \approx 2\.8\pm0\.3$ [1$\sigma$ error, from
their Fig.~3(a)].

This is close to our best estimate for $\Omega_0=1, \lambda_0=0,$
but lower by $0\.4${}$\sigma$ than our result for $\Omega_0=0\.1, \lambda_0=0$
(\S\ref{s-r0res}, using our uncertainty). So even without knowing 
what value of $\Omega_0$ corresponds to their best estimate for $r_0,$ 
their result is consistent with ours. The caveats regarding the 
use of eq.~(\ref{e-ls93ic}) and not correcting for regions biased by
large objects also apply here.

Other correlation function estimates based on either spectroscopic 
or photometric redshifts include those of \citet{CFRS-VIII} 
(spectroscopic, \zmed$=0\.53$)
and \citet{GiavDCP98} (spectroscopic, \zmed$=3\.0$) and 
\citet{MPR99} (photometric, \zmed$=3\.7$). 

The first two of these estimates are for fields of angles about a
factor of ten larger than that of an WFC field, so cover the transition
scale between the quasi-linear and non-linear regimes of galaxy clustering and 
may not express clustering fixed in proper units. Nevertheless,
we discuss these briefly.

\citeauthor{CFRS-VIII}'s \zmed$=0\.53$ result for the present-day value of 
$r_0$ and $\epsilon=0$ (\S{4.1-1.} of \citealt{CFRS-VIII}) is 
$r_0=3\.0\pm0\.2${\hMpc} ($\Omega_0=1, \lambda_0=0$) 
or $r_0=3\.9\pm0\.2${\hMpc} ($\Omega_0=0\.2, \lambda_0=0$). 
This is consistent with our result.
However, this is likely to be 
a coincidence. The effect of a non-zero size of $\rhalo$ 
which can decrease the inferred $r_0$ on the HDF scale studied by 
\citeauthor{Vill97}, \citeauthor{Conn98} and ourselves, is
replaced in the case of \citeauthor{CFRS-VIII}, by clustering
growth stronger than $\epsilon=0$  which decreases the values
of $r_0$ at low redshifts (cf. \S\ref{s-dcp}).

It should be kept in mind that the galaxies in our sample 
\citep{Mob98}
are selected at rest wavelengths of around $1000$\AA, so 
comparison with correlation functions estimated for 
low redshift galaxies is difficult. \citeauthor{Mob98} find that 
the galaxies in the sample are mostly ellipticals and starburst
galaxies. 

At low redshifts, the former have slightly higher
correlations than the general galaxy population. 

If the 
latter are a random selection of young disk galaxies which have
just collapsed and started a rapid burst of star formation, 
then the corresponding correlation implied from low redshifts
should be slightly lower than that of the total population.
On the other hand, if only those young disk galaxies which happen
to be close to one another have star bursts in order to be
selected among our HDF `UV drop-in' sample, then the correlation
could well be higher.

Modelling the combination of the two populations would also require
knowledge of the cross-correlation between them. 

So, the overall population mix observed at $z\sim2$ is likely to 
represent a complex mix of the populations at low redshifts.
The different effects might cancel each other out, or cause
a significant systematic effect.

\subsection{The Decreasing Correlation Period} \label{s-dcp}

\citeauthor{GiavDCP98}'s result at \zmed$=3\.0$ for Lyman break galaxies 
(LBG's) is for a large value of $r_0.$
Expressed as values for clustering fixed in proper
units
[i.e. multiplying by $(1+$\zmed$)^{[0-(\gamma-3)]/\gamma}$]
this is 
$r_0=5\.3^{+1.0}_{-1.3}${\hMpc} ($\Omega_0=1, \lambda_0=0$) 
or $r_0=8\.4^{+1.8}_{-1.5}${\hMpc} ($\Omega_0=0\.2, \lambda_0=0$). 
This is about $2${}$\sigma$ larger than our estimate.

As these authors discuss, this seems to be a detection of the
decreasing correlation period 
(DCP; \citealt{Rouk93,BV94b,ORY97,Bagla98,SteiDCP98,Mosc98,Coles98}). 
The DCP is the period
of large galaxy formation when the transition from linear perturbations
to non-linear collapsed haloes may have caused a high bias factor 
in the initial correlation function of these haloes, 
which later disappeared as perturbations in underdense regions also
collapsed. 

Since most of the signal in $w$ of \citeauthor{GiavDCP98} is for
$\theta > 20\arcs,$ this is corresponds mostly to larger length scales
than our measurements. Their point which appears to be for 
a 0-20{\arcs} bin (Fig.~2 their paper) is lower than their power law
fit, so they may not have a detection of the DCP at this scale.

Nevertheless, the DCP is clearly absent at the small scales measured
in this paper from the HDF. 
Following \citet{ORY97}, 
a simple power law extension of eq.~(\ref{e-xieps})
is 
\begin{equation}
\xi(r,z)=\left\{ 
        \begin{array}{ll}
        \left[(1+z)\over(1+z_t)\right]^\nu \xi(r,z_t),
                & z > z_t \\
        (r_0/r)^\gamma (1+z)^{-(3+\epsilon-\gamma)}, & z_t \ge z >0
        \end{array}
        \right.
\label{e-eps_nu}
\end{equation}
where correlation evolution and parameters for low redshift 
are as for eq.~(\ref{e-xieps}), but a transition redshift $z_t$  
and DCP slope $\nu$ parametrise the evolution during the DCP.

If the present study does not miss the DCP due to the smallness of 
the length scale studied, then our result combined with that of 
\citeauthor{GiavDCP98} would imply that $z_t$ lies
somewhere in the range $2 < z_t < 3.$ 

This would imply that the
transition from the stage of the first collapse of large haloes
in high density regions to the stage when most haloes of the same
mass scale have collapsed finishes just before the time of
major star formation (cf. \citealt{Madau96}). Corrections 
to \citeauthor{Madau96}'s plot for dust
(e.g. \citealt{Mob98,Hughes98}) 
would imply an earlier and less sharply peaked maximum in 
the volume-averaged star formation history, so the overlap between
the two epochs would seem to be even stronger.
That is, the major star formation period would
start just when most haloes of large mass have collapsed. 

Finally, we consider the analysis of \citet{MPR99} for HDF galaxies
at \zmed$= 3\.73.$ The measured correlation function for these
galaxies is less easy to interpret than that of the present 
paper. The most significant correlation is in the bin centred
at $\theta\approx 30\arcs$. This could 
either be (a) a clustered region 
occurring by chance at this particular redshift, or (b) a detection 
of the DCP to a slightly higher redshift than that of
\citeauthor{GiavDCP98}. Expressed 
for an $\Omega=1,\lambda=0$ universe and $\epsilon=0$, 
the correlation length is $r_0=7\.1\pm1.5${\hMpc} \citep{MPR99}. 

That is, the amplitude is higher than that of \citet{GiavDCP98},
expressed in terms of the same parameters. Since it is at a higher 
redshift this is just what would be expected from the DCP.

If we take the two estimates together as estimates of the
initially highly biased correlation above that expected for clustering fixed
in proper coordinates, 
adopt 1$\sigma$ 
uncertainties of $\delta($\zmed$)=0\.1$ for the two data sets,
and use $1-\gamma=-0\.8,$
then the value of $\nu$ in eq.~(\ref{e-eps_nu}) is
$\nu = 2\.1\pm3\.6.$ Moreover, if we 
consider our result and that of \citet{Conn98} 
to be the `stable' non-linear correlation length achieved at
the end of the DCP, $r_0 \sim 2\.6\pm1\.4,$ then the transition redshift
marking the end of the DCP is $z_t = 1\.7\pm0\.9.$ 

This is clearly consistent (within the error bars) with the lack
of detection of the DCP in the present work. That is, the extrapolation
of a simple power law evolution through the estimates of 
\citet{MPR99} and of \citet{GiavDCP98} implies that the DCP finishes
just slightly below $z=2,$ i.e. has mostly disappeared by the epoch
of our estimate.

In addition, these values happen to lie well within the constraints
derived from diverse $N$-body models (\S3, \S6, \citealt{ORY97}).

However, since the \citeauthor{GiavDCP98} result is for larger
scales, it would be more consistent to compare the two at scales 
for which stable clustering in proper units may not be
valid, in which case the analysis would be a lot less trivial than
the simple estimate made here.

\subsection{Halo Radii}

Our value of $r_0=2\.6^{+1.1}_{-1.7}${\hMpc} is lower than typical
low redshift values of $r_0\sim 5${\hMpc}. As shown in 
\S\ref{s-rhalocut} and \S\ref{s-angdep}, for $r_0=5\.5${\hMpc} and
$\epsilon=0$, the correction for halo pair exclusion
implies very reasonable values of $\rhalo$ to 
match the measured value of $w(5\arcs).$  

Matching the
\mbox{12-21{\arcs}} bin in addition (if ignoring systematic uncertainties),
would require a low density 
metric, a higher value of $\epsilon$ or a moderate anti-bias,
in which case the correction for non-zero `halo radii' is 
likely to be considerably smaller for the present data set. 

Could the effect of $\rhalo$ be detected in other data sets ? 
Both Fig.~5 of \citet{MPR99} and Fig.~3 of \citet{Post98}
show decreases in slope below around 30{\arcs} and 60{\arcs} respectively,
which are qualitatively similar to those in Fig.~\ref{f-wthprop} and
Fig.~\ref{f-wthcomov}. 

In Fig.~\ref{f-wthprop}, the effect of $\rhalo=100${\hkpc} 
could be expected to
occur below angles corresponding to perpendicular (proper) distances
of 100{\hkpc}, i.e. 14-24{\arcs} for the three sets of metric parameters.
It is clear that the effect, integrated precisely via Limber's equation, 
occurs just slightly below these angles. 

In that case we can estimate the values of $\rhalo$ 
below which values of $w$ falling below 
a projected $1-\gamma=-0\.8$ power law could be explainable simply 
by our halo cutoff formula. The angles in 
Fig.~5 of \citet{MPR99} and Fig.~3 of \citet{Post98} at which these
falloffs occur (in the latter just the two fainter slices are considered)
would imply values of $\rhalo$ just slightly larger than 
100-210{\hkpc} and 250-360{\hkpc} respectively. 

The former is reasonable. The latter is rather large, and for 
`halo exclusion' to apply, would require an explanation of why 
galaxies rarely occur close to each other (relative to the expectation
from a power law $\xi$) at $z\sim 0\.8$, but are able to occur this
closely for long enough periods at lower redshifts, such that the 
effect disappears in \citeauthor{Post98}'s analyses for brighter
magnitudes.  

Low redshift estimates of $\xi$ at short separations would be good
for comparison, but few exist. 
\citeauthor{DavP83} (1983) analysis is consistent with 
eq.~(\ref{e-xi}) on scales 
$10${\hkpc}$ \ltapprox r \ltapprox 10${\hMpc}, but the correlations
are quite noisy at the small scales of interest.
A more recent estimate is that of \citet{Tuck97}, who 
finds a similar result down to about $20${\hkpc}, but 
for the redshift-space correlation function rather than the `real'
(i.e. spatial) correlation function, so this also 
is not easy to interpret.

\section{Conclusions}\label {s-conclu}

We have estimated 
the amplitude of the angular two-point galaxy auto-correlation
function $w(\theta)$ for galaxies at $z\sim 2$ from a
a $U<27$ complete sub-sample of the HDF, and find 
a result compatible, though slightly higher, than that of 
\citet{Vill97}, although the two samples 
have similar estimated median redshifts. The use of photometric
redshifts, the avoidance of linear error terms in the estimator
used for calculating the correlation function and the masking 
of regions biased by large objects favour our result as the
more accurate of the two estimates; but the smallness of the
numbers of galaxies in our sample favours that of \citeauthor{Vill97}
as the more precise.

The consistency between the two samples suggests that the high star
formation rate and domination by starburst galaxies and ellipticals in
our $U$-band selected sample compensates for any effects due to the
difference in rest-frame wave-bands between our sample and other
samples. Since $U$-band selection also has the advantage
of a good high redshift cut-off and helps to estimate photometric redshifts,
this seems a useful strategy to complement selection in other wavebands.
The technique could be referred to as the `UV drop-in' technique 
(cf. \citealt{Stei96}).

(i) Use of eq.~(\ref{e-wham24}), i.e. eq.~(24) of \citet{Ham93}, is
illustrated in observational data, possibly for the first time.
It is shown how this compares to the use of eq.~(\ref{e-ls93ic}).
Both equations are similar to that of \citealt{LS93}, but 
eq.~(\ref{e-wham24}) 
corrects for uncertainty in the mean number density, 
$\nmean,$ without
re-introducing linear error terms.

The estimate using eq.~(\ref{e-wham24}) requires an external
estimate of $\nmean,$ which we estimate from the mean of the three
WFC fields, considering the individual fields as independent
experiments (but with identical selection criteria).
This gives $w(\theta\approx5\arcs) =0\.10 \pm 0\.09.$
Alternatively, adoption of the conventional constraint 
that $w$ should be a power law of a given slope, for 
$1-\gamma=-0\.8,$ increases this by 60\%, but the 
desired slope cannot be exactly obtained. Given the former
estimate and its uncertainty, 
the possibility that the latter is identical to 
the former is only rejected at a 50\% confidence level. Since
the variance is related to the uncertainty, it is unsurprising
that the difference is not formally of high significance.

Eq.~(\ref{e-ls93ic}), which requires an assumption about the
shape and slope of $w,$ and which is normally sure of 
obtaining this slope, 
implies values about twice as high as these, 
depending on what value of $1-\gamma$ is assumed. 
This can be rejected at the 68\% confidence level given 
a true value and error distribution as estimated 
using eq.~(\ref{e-wham24}) and $\nmean$ from the mean of the 
three fields. However, relative to the 
estimate from constraining  eq.~(\ref{e-wham24}) to a 
power law  of the same slope, this is only rejected at the 24\%
confidence level.
Again, the differences are related to the uncertainty terms, so
are of a similar order of magnitude.

(ii) Biases introduced in faint galaxy selection by large
objects, in this case by what appears to be the `hiding' of galaxies
by Poisson noise, creating `voids', increases the estimate cited
by 60\% if it is not corrected.  The null hypothesis of not masking
introducing no change is rejected only at the 50\% confidence
level.

The difference in estimates of $w$ with and without
merging together of objects closer to one another than $0\.5\arcs$ 
is negligible at 5{\arcs}, though is noticeable at larger angles.
This favours (marginally) 
the `{\HII} region' interpretation of sub-arcsecond HDF galaxy
pairing relative to the `building blocks' interpretation 
(see \citealt{Colley96,Colley97}).

(iii) The scales effectively studied here 
are in the range $\approx 25${\hkpc}$-250${\hkpc}, 
so overlapping of gaseous and/or dark matter haloes 
should exclude some fraction of galaxy pairs. 
A simple formalism for correcting $\xi$ by a smooth (Gaussian)
cutoff [eq.~(\ref{e-xirhalo})] is presented.
For clustering stable in proper units [$\epsilon=0$ in 
eq.~(\ref{e-xieps})]
in an $\Omega=1,\lambda=0$ universe, 
our $w(5\arcs)$ estimate 
(a) implies a present-day correlation length of 
$r_0\sim2\.6^{+1.1}_{-1.7}${\hMpc} if halo 
sizes are ignored, but (b) for 
a present-day correlation length of $r_0=5\.5${\hMpc} 
implies that a typical halo radius is
$\rhalo=70^{+420}_{-30}$\hkpc. 

This value of $\rhalo$ is comfortably close 
to what could be expected, although 
this correction was not devised to fit the present data. It
was pointed out by \citet{RV97} as simply being a correction likely
to be needed just because halo and galaxy sizes are nonzero.

However, comparison of $w(\theta)$ as integrated using Limber's
equation with that estimated for the $z\sim2$ galaxies shows that this
correction is insufficient to bring an $\Omega=1,\lambda=0$ universe
into agreement with the data for clustering stable in proper
units.  Indeed, the second bin has a more significant estimate
of $w$ than the first, if only statistical uncertainties are 
considered: $w(\approx 16\arcs)= 0\.063\pm0\.018.$ 
However, comparison of  Figs~\ref{f-wiform2}-\ref{f-wiform1} shows
that systematic uncertainties are probably a few times larger than
the statistical uncertainties for this (and larger) bins.

In the case of a cosmological constant dominated universe
($\Omega_0=0\.1, \lambda_0=0\.9$), $\rhalo$ has little effect, and as
is already well known for previous estimates of $w$
(e.g. \citealt{YPT93}), a good fit over all angles results for
low density metrics and $\epsilon=0$, 
with or without a cosmological constant to flatten the metric. 
Higher values of $\epsilon$ or anti-bias should have
similar effects.
However, the systematic uncertainties in the larger angular bins
imply that a detailed fit
across all angles may not be justified.

It should also be noted that the lack of an internal correction for
sample variance is a common feature of all the HDF angular 
correlation function estimates, because of the HDF's small size.
This introduces uncertainty in all of these estimates.
For the lowest redshift samples, a power law constraint based on
large solid angle, 
faint apparent magnitude/surface brightness limited samples 
could provide an observationally justified way to reduce this
error. For the objects at successively higher redshifts, 
surveys using 
UV drop-in and UV drop-out techniques over large solid angles 
could provide observational estimates of the large scale variance.

(iv) The results of \citet{GiavDCP98} and \citet{MPR99} 
can be expressed, for an $\Omega=1,\lambda=0$ universe and $\epsilon=0$, 
as zero-redshift correlation lengths of $r_0=5\.3^{+1.0}_{-1.3}${\hMpc} and 
$r_0=7\.1\pm1\.5${\hMpc} respectively.
These are both strongly suggestive of a decreasing correlation 
period (DCP; \citealt{Rouk93,BV94b,ORY97,Bagla98,SteiDCP98,Mosc98,Coles98}),
during which the first haloes to collapse do so in high 
density regions, so are highly biased relative to the underlying
density perturbations. The period terminates when most
haloes of the same mass scale, in both high and low density regions,
have collapsed, so that the correlation function behaves `normally',
i.e. at small scales is stable in proper coordinates.

The decreasing correlation period (DCP) of a high initial bias
in the spatial correlation function is not detected in our data,
so would have to terminate in the range 
$2 \ltapprox z_t \ltapprox 3.$ 

A simple fit to the two values 
just cited for \citeauthor{GiavDCP98} and \citeauthor{MPR99}
would imply values of 
$z_t = 1\.7\pm0\.9$ and $\nu=2.1\pm3.6$ in \citeauthor{ORY97}'s
extension of the standard power law fit [eq.~(\ref{e-eps_nu})]. 
That is, values of $\xi$ at redshifts greater than 
$z_t = 1\.7\pm0\.9$ would be $[(1+z)/(1+z_t)]^{2.1\pm3.6}$ times their 
values at $z_t$, for a fixed value of $r$ in comoving units.

These values are consistent with those estimated from $N$-body
simulations (\citealt{ORY97}) and 
with our lack of detection of the DCP.

However, since the three studies are optimised to different scales,
and since \citeauthor{MPR99}'s result might be that of an
individual structure which is not statistically representative,
this parametrisation of the DCP should be taken with caution.

\section*{Acknowledgements}
This research has been supported by the 
Polish Council for Scientific Research Grant
KBN 2 P03D 008 13 and has benefited from 
the Programme jumelage 16 astronomie 
France/Pologne (CNRS/PAN) of the Minist\`ere de la recherche et
de la technologie (France).
Use has been made of the HDF archive of the Space Telescope Science 
Institute at 
{\em http://www.stsci.edu/ftp/science/hdf/archive/v2.html} and
of the resources of the 
Centre de donn\'ees astronomiques de Strasbourg (CDS).


\end{document}